\documentclass[prd,preprintnumbers,
twocolumn,
eqsecnum,floatfix,letterpaper,superscriptaddress,nofootinbib]{revtex4-1}
\usepackage[usenames,dvipsnames]{color}
\usepackage{amsmath,amssymb,graphicx}
\usepackage{tensor}
\usepackage{bm}
\usepackage{times}
\usepackage{microtype}
\usepackage{booktabs}
\usepackage{subfigure}
\usepackage[normalem]{ulem}
\usepackage{todonotes}
\usepackage[varg]{txfonts}
\usepackage[colorlinks, pdfborder={0 0 0}]{hyperref}
\usepackage{xfrac} 
\definecolor{LinkColor}{rgb}{0.75 , 0, 0}
\definecolor{CiteColor}{rgb}{0, 0.5, 0.5}
\definecolor{UrlColor}{rgb}{0, 0, 0.75}
\hypersetup{linkcolor=LinkColor}
\hypersetup{citecolor=CiteColor}
\hypersetup{urlcolor=UrlColor}
\allowdisplaybreaks
\usepackage[CaptionAfterwards]{fltpage}

\usepackage[section]{placeins}
\usepackage{multirow}

\usepackage{array}
\newcolumntype{L}[1]{>{\raggedright\let\newline\\\arraybackslash\hspace{0pt}}m{#1}}
\newcolumntype{C}[1]{>{\centering\let\newline\\\arraybackslash\hspace{0pt}}m{#1}}
\newcolumntype{R}[1]{>{\raggedleft\let\newline\\\arraybackslash\hspace{0pt}}m{#1}}

\begin{document}
\newcommand{\FigStart}{\begin{figure}[h]}
\newcommand{\blue}{\color{blue}}
\newcommand{\red}{\color{red}}
\newcommand{\msun}{$M_{\odot}$} 
\newcommand{\Msun}{M_{\odot}} 
\newcommand{\vek}[1]{\boldsymbol{#1}}
\newcommand{\CMI}{Chennai Mathematical Institute, Plot H1, SIPCOT IT Park, Siruseri, 603103 Tamilnadu, India}
\newcommand{\IGC}{Institute for Gravitation and the Cosmos, The Pennsylvania State University, University Park, USA}
\newcommand{\AEI}{Max Planck Institute for Gravitational Physics (Albert Einstein Institute), D-14476 Potsdam-Golm, Germany}
\newcommand{\PSUAstro}{Department of Astronomy and Astrophysics, The Pennsylvania State University, University Park PA USA}
\newcommand{\PSUPhys}{Department of Physics, The Pennsylvania State University, University Park PA USA}
\newcommand{\CFPhys}{School of Physics and Astronomy, Cardiff University, Cardiff, CF24 3AA, United Kingdom}
\newcommand{\I}{\mathrm{i}}
\newcommand{\E}{\mathrm{e}}
\newcommand{\eff}{\text{eff}}
\newcommand{\Tc}{\, \text{,}}
\newcommand{\Td}{\, \text{.}}
\newcommand{\md}{\mathrm{d}}

\newcommand{\numberthis}{\stepcounter{equation}\tag{\theequation}} 

\title{Comparison of post-Newtonian Mode Amplitudes with Numerical Relativity Simulations of Binary Black Holes}

\author{S.~Borhanian}
\affiliation{\IGC}
\affiliation{\PSUPhys}
\email{sub284@psu.edu}
\author{K.~G.~Arun}
\affiliation{\IGC}
\affiliation{\CMI}
\email{kgarun@cmi.ac.in}
\author{H.~P.~Pfeiffer}
\affiliation{\AEI}
\email{harald.pfeiffer@aei.mpg.de}
\author{B.~S.~Sathyaprakash}
\affiliation{\IGC}
\affiliation{\PSUPhys}
\affiliation{\PSUAstro}
\affiliation{\CFPhys}
\email{bss25@psu.edu}
\date{\today}
\begin{abstract}
Gravitational waves from the coalescence of two black holes carry the signature of the
strong field dynamics of binary black holes. In this work we have used
numerical relativity simulations and post-Newtonian theory to investigate this
dynamics.  Post-Newtonian theory is a low-velocity expansion that assumes the
companion bodies to be point-particles, while numerical relativity treats black
holes as extended objects with horizons and fully captures their dynamics.
There is {\em a priori} no reason for the waveforms computed using these
disparate methods to agree with each other, especially at late times when the black
holes move close to the speed of light. We find, remarkably, that the leading
order amplitudes in post-Newtonian theory agree well with the full general
relativity solution for a large set of spherical harmonic modes, even in the
most dynamical part of the binary evolution, with only some modes showing
distinctly different behavior than that found by numerical relativity
simulations. In particular, modes with spherical harmonic indices $\ell=m$ as
well as $\ell=2,m=1$ are least modified from their dominant post-Newtonian
behavior. Understanding the nature of these modes in terms of the post-Newtonian
description will aid in formulating better models of the emitted waveforms in
the strong field regime of the dynamics.
\end{abstract}
\maketitle
\section{Introduction and motivation}\label{sec:Intro}
The Laser Interferometer Gravitational-Wave Observatory (LIGO) at two sites in
the USA (Hanford, WA and Livingston, LA) and the Virgo detector in Pisa, Italy,
have opened a new era in multi-messenger astronomy and fundamental physics via
the discovery of binary black
hole~\cite{GW150914,GW151226,GW170104,GW170608,GW170814} and binary neutron
star~\cite{GW170817} mergers. These discoveries have, for the first time,
enabled tests of dynamical gravity in the strongly dissipative regime of the
theory~\cite{TOG,O1BBH,GW170104,GW170814}, i.e. the period derivative of the
binary $\dot{P}$ changes very rapidly during the time of observation (see
Refs.~\cite{WillLR05,SathyaSchutzLivRev09,YunesSiemens2013,GairLivRev}). This is
in contrast to the Hulse-Taylor binary~\cite{HulseTaylor75} where the change in
period $\dot P$ is essentially constant.

Radio measurements of the rate at which the orbital period decays in a binary
neutron star allowed spectacular confirmation of the quadrupole
formula~\cite{TFMc79,TW82,Lyne:2004cj,Kramer:2005ez}. However, radio binary
pulsars probe the weak field sector of the two-body dynamics\footnote{We note
that the self-gravity of the neutron stars, which must be taken into account in
the measurement of the various binary parameters, are large. Indeed, the
compactness of neutron stars given by the dimensionless quantity ${\cal C}
\equiv GM_{\rm NS}/c^2R_{\rm NS},$ where $M_{\rm NS}$ is the mass of the neutron
star and $R_{\rm NS}$ its radius, is about ${\cal C} \sim 0.2.$ In this sense, the observations
do probe the strong field regime of general relativity; however, the two-body dynamics
is governed by weak fields.}, wherein the dimensionless gravitational potential
$\phi$ of one of the bodies on the other is $\phi \ll 1,$ or, equivalently, the
speed $v$ obeys $v/c \sim \sqrt{\phi} \ll 1.$ In contrast, gravitational wave observations of
the merger make it possible to test general relativity when $\phi \sim 0.5$ (the largest it ever
gets) and the system is strongly dissipative.  Consequently, LIGO, Virgo, and
other future ground-based gravitational wave detectors (KAGRA and LIGO-India) can test the validity of
general relativity in an entirely new regime of the theory. 

\subsection{Modeling binary black hole dynamics} 
The dynamics of a binary black hole consists of three phases: inspiral, merger,
and ringdown. Inspiral refers to the early phase of the binary evolution when
the effect of radiation reaction on the orbital motion is small. The
slow-motion, weak-field dynamics of this phase, when the two black holes are far
apart, is well-modeled by post-Newtonian (PN) theory (see Ref.~\cite{Bliving}
for a review)  where all the observables are expressed as a power series in
$v/c$. The strong field dynamics close to the merger and the dynamics of the
highly deformed remnant black hole can only be modeled using numerical
relativity, where one solves Einstein's equation for the two-body
problem using numerical techniques (see Ref.~\cite{Pretorius07Review} for a
review). The ringdown phase of the dynamics occurs when the remnant black hole
has become less deformed and can  be well-approximated as a perturbation of a
Kerr black hole and modeled using black hole perturbation theory (see
Ref.~\cite{TSLivRev03} for a review). 

The waveform emitted by an inspiralling compact binary predominantly consists
of the quadrupole mode. It was pointed out that controlling the evolution of
the orbital phase of the dominant mode was far more important \cite{Cutler:1992tc} for the 
detection problem than controlling the correction to its amplitude or the inclusion
of higher order modes that contain wave frequencies other than
twice the orbital frequency of the quadrupole mode.  In this so-called {\em restricted
post-Newtonian approximation} one neglects the correction to the amplitude of the
waveform arising from higher order multipoles. However, higher modes are critical
for an unbiased estimation of both the intrinsic parameters of a binary (e.g. 
companion masses and spins) but also the orientation of the binary relative to
a detector and its position on the sky \cite{VanDenBroeck:2006qi,VanDenBroeck:2006qu,
Arun:2007hu,AISSV07,AMVISS09}. Waveforms based on this new information could
be useful to test general relativity in the high-curvature regime of the theory, which
is one of the principal goals of gravitational wave astronomy.

Understanding the structure of the multipole modes and their dependence on the
intrinsic parameters of the binary will be critical in building more refined
waveform models and using gravitational wave observations to test predictions
of general relativity in dynamical spacetimes.

\subsection{Imprints of progenitors on the black hole ringdown spectrum: Past
studies}

Previous studies \cite{Kamaretsos2012a, Kamaretsos2012b} found that
the properties of the  progenitor system such as the symmetric
mass ratio $\eta=m_1m_2/(m_1+m_2)^2$ (where $m_{1,2}$ denote the
binary component masses)  or mass ratio-weighted combinations of component spins
(referred to as ``effective spin" parameter)
leave their unique imprints on various ringdown modes of the remnant
black hole. More specifically,
Ref.~\cite{Kamaretsos2012a} had found that for mergers of nonspinning
black holes, the amplitudes of the four strongest modes, (2,2), (2,1),
(3,3), (4,4), continue to depend on mass ratio even $15M$ after the luminosity of the (2,2) mode peaks.
In a follow-up work, Ref.\,\citep{Kamaretsos2012b}
further investigated binaries whose component spins are
aligned with the orbital angular momentum. The authors found that
the relative amplitude of the (2,1) mode $10M$ after the peak luminosity could be 
captured by a fitting formula with two variables: $\eta$ and an effective spin 
parameter. 

While the fitting function for the (2,2) mode in
Ref.\,\cite{Kamaretsos2012a} was based on physical intuition gained
from PN theory, the ansatz for the other modes were based on fits to the numerical data.
Similarly, \cite{Kamaretsos2012b} uses a functional form for the
(2,1) mode inspired by PN theory including not only
symmetric mass ratio, but also spin dependencies.
Achieving a good fit required an effective spin combination that was slightly
different from the one found in PN approximation \citep{Mishra:2016whh}.
These results pointed to the interesting possibility of inferring
the properties of the progenitor black holes just from the late ringdown
signal.

Following a different approach \cite{London:2014cma}, London et
al.~studied the $\eta$-dependence of the higher modes of the
post-merger amplitudes. Their study is based on fitting the amplitudes
of higher modes from numerical relativity simulations to high-order polynomials
in the symmetric mass ratio. Such fits are useful in building
analytical models of the post-merger waveforms.  Indeed, in a more
recent study~ \cite{London2018}, London et al.~developed a new
phenomenological waveform model that includes higher modes. 
\subsection{Present work}
In the present work we use a combination of numerical relativity
simulations and PN theory to study the evolution of
different modes of gravitational waves as a function of time, mass
ratio and mode-dependent ``effective spin" parameters (see Eq. 
\ref{eq:effective_PN_spins}).  The phase evolution 
of each mode, being a multiple of the orbital phase, is essentially the 
same for all modes and has been amply treated in the literature; we, 
therefore, restrict our study entirely to the mode amplitudes.
The two gravitational wave polarisations $h_+$ and $h_\times$
from an inspiralling binary are, in principle, composed of infinitely many
modes as exemplified by the relation \cite{Th80}
\begin{equation}
    h_+-i\,h_{\times}
    =\sum_{\ell=2}^{\infty}\sum_{m=-\ell}^{\ell}\,h_{\ell m}\,_{-2}Y_{\ell m}(\theta,\phi) \Tc \label{eq:polarisations_and_modes}
\end{equation} 
where $_{-2}Y_{\ell m}$ are the $-2$ spin-weighted spherical harmonics,
$(\theta,\phi)$ define the direction of propagation of the wave, and
$h_{\ell m}$ are the spherical harmonic wave modes. Although the
quadrupole $(\ell,m)=(2,2)$ is the dominant mode, higher order modes
can have comparable, albeit smaller, amplitudes relative to the
quadrupole when the component masses are very different or the compact
objects have significant spin. In the inspiral regime, using
a cocktail of approximation schemes,
PN theory provides an effective framework to relate the
{\it radiative multipoles} observed at infinity to  the {\it source
multipoles}~\cite{Bliving,BDI95}, thereby expressing the observed
gravitational waveform in terms of the multipole moments of the source.

Kelly and Baker \cite{Kelly:2012nd} investigated the effects of mode mixing
between the spherical harmonics, used in numerical relativity and PN theory, and spheroidal
harmonics. The latter capture the axial symmetry of the Kerr spacetime and
hence are a more suitable basis to describe perturbations of the Kerr metric
during the ringdown. In particular, they showed that the $(3,2)$ spherical
harmonic mode has significant contributions from different spheroidal harmonic
modes, which is referred to as \emph{mode mixing}. A spheroidal harmonic
decomposition renders the modes to fall off more smoothly as a function of
time, thus allowing a more simplified modeling of the waveform.  While this is
true, we provide an alternative interpretation of mode mixing as arising due to
the failure of the point particle description of PN theory close to the
formation of a common horizon.

This formation marks a rather intriguing transition in the binary black
hole dynamics: from the perturbative dynamics of the two black holes to the
perturbative dynamics of the remnant black hole via this highly
non-perturbative merger. This transition is captured in the full general relativity solution,
provided by numerical relativity simulations that by design track the dynamics of black hole
horizons. Hence, given the availability of numerical relativity catalogs for binary black hole
mergers, it is interesting to ask how the information about the two black
holes, encoded in PN expressions, propagates from the inspiral to the merger
and ringdown phases and whether with just observing the latter two, one can
infer the properties of the binary components.  Here, extending the works of
\cite{Kamaretsos2012a,Kamaretsos2012b}, we compare several
spherical harmonic mode amplitudes from SXS numerical
simulations~\cite{SXSWebsite} with the leading terms in the corresponding PN
expressions; allowing for one free parameter in the nonspinning case and two
free parameters in the spinning case. Our aim is to search for those modes
that are fitted very well by the aforementioned PN-based fits, and hence
retain information about the progenitor system. 

We find that the most dominant mode amplitudes hardly change their dependence on the 
symmetric mass ratio, given from PN theory, throughout the evolution of the 
binary.
Further, we find that the signature of the strong field regime
is encoded in a small number of modes that are sub-dominant, with their amplitude being 
less than 10\% of that of the quadrupole.

The paper is organized as follows: Section~\ref{sec:SXS}
describes the
SXS numerical simulations we employ for the study. The fitting model we
use, based on PN expressions for leading order spherical harmonic
modes of the waveform, is explained in Sec.~\ref{sec:PN}. Our results on
the PN signatures in the spherical harmonic modes of numerical
relativity are described in Sec.~\ref{sec:PNSig} and the implications of
these results for modelling waveforms from binary black holes are discussed in
Sec.~\ref{sec:Implications}. Appendix~\ref{app:alternative_mass_ratio_series} provides an alternative representation of some rmesults in Section~\ref{sec:PNSig}. Some of the technical details of the
simulations are elaborated in appendix~\ref{app:SXS_simulations}.

\section{Numerical simulations}\label{sec:SXS}

This study utilizes publicly available binary black hole gravitational waveforms
from the SXS collaboration~\cite{SXSWebsite}. The concrete
simulations used are listed in Appendix~\ref{app:SXS_simulations}.
Specifically, Table~\ref{tab:simulations_ns} lists the 43 non-spinning
simulations that were used, while Table \ref{tab:simulations_as} lists
the 121 aligned-spin simulations.  
The simulations were originally presented as follows:
\begin{itemize}
\item The first SXS waveform catalog~\cite{Mroue:2013xna} ($1\le$ SXS id $\le 174$).
 \item Simulations for developing techniques for very high black hole spins~\cite{Lovelace:2014twa,Scheel:2014ina}  ($175\le$ SXS id $\le 178$).
 \item Simulations for a waveform surrogate model for non-spinning binary black hole systems~\cite{Blackman:2015pia}  $180\le$ SXS id $\le 201$).
 \item Binary black hole simulations at mass-ratio 7 with particularly many inspiral cycles~\cite{Kumar:2015tha} ($202\le$ SXS id $\le 207$).
\item A study of aligned spin binary black hole systems~\cite{Chu:2015kft,Kumar:2016dhh} ($209\le$ SXS id $\le 304$).
\end{itemize}

The simulations were computed with the Spectral Einstein Code
(\texttt{SpEC})~\cite{SpECwebsite}, a multi-domain pseudo-spectral
code designed to solve elliptic and hyperbolic partial differential
equations, in particular the Einstein equations.  \texttt{SpEC}
computes initial data with the extended conformal thin sandwich
method~\cite{York1999,Pfeiffer2003b} utilizing quasi-equilibrium black hole
excision boundary conditions~\cite{Cook:2004kt, Caudill:2006hw,
  Lovelace:2008tw} and iterative eccentricity
reduction~\cite{Buonanno:2010yk} to achieve quasi-circular inspirals.
\texttt{SpEC} evolves the Generalized Harmonic form of Einstein's
equations~\cite{Friedrich1985,Pretorius:2004jg} in first order
form~\cite{Lindblom:2005qh} with constraint
damping~\cite{Gundlach:2005eh, Pretorius:2004jg, Lindblom:2005qh} and
constraint preserving boundary conditions~\cite{Lindblom:2005qh,
  Rinne:2006vv, Rinne:2007ui}.  The code uses black hole excision~\cite{Scheel:2008rj, Szilagyi:2009qz,
  Hemberger:2012jz}, coupled with a dual-frame approach to have the
computational grid track the motion of the black hole
horizons~\cite{Scheel:2006gg}.  The gravitational wave data used in
our study was extracted with Regge-Wheeler-Zerilli
wave-extraction~\cite{Sarbach:2001qq,Regge:1957td, Zerilli:1970se}, and corrected for time-dilation effects at the extraction radius~\cite{Boyle:2009vi,Taylor:2013zia} and for mode-mixing arising from small residual motion of
the center of mass~\cite{Boyle:2015nqa}.
More technical details are given in the original publication
presenting the
simulations~\cite{Mroue:2013xna,Lovelace:2014twa,Scheel:2014ina,Blackman:2015pia,Kumar:2015tha,Chu:2015kft,Kumar:2016dhh}.

\texttt{SpEC} simulations are generally run at multiple numerical
resolutions, in order to be able to assess numerical convergence and numerical truncation
error. Indeed, we have restricted the present study only to
simulations that are available at multiple resolutions.  The last
column in Tables~\ref{tab:simulations_ns} and~\ref{tab:simulations_as}
lists the resolutions of each simulations that were used.  For each
simulation, the accuracy increases with a larger numerical value in
this column.  However, because of improvements to SpEC's numerical
algorithms in the course of time, it is not possible to assign an absolute
meaning to these resolution values.  Using the different numerial
resolutions, we compute an error bar for every numerical value
extracted from the numerical relativity data based on the difference in this
value when extracted from the numerical relativity data of different resolution.

Visual inspection of the $(\ell,m)$ modes indicate that the $(2,2)$ mode is
well-behaved for the simulations considered here. The leading sub-dominant
modes with $m\le 4$, specifically (2,1), (3,3), (3,2), (3,1), (4,4), (4,3),
(4,2), (4,1), are also generally well-behaved with only rare visible unphysical
features, like for instance unexpected extraneous oscillations during ringdown.
Further, the numerical errors of these modes, see Figs.
\ref{fig:non_spinning_plots} and \ref{fig:aligned_plots_PN_time_series},
indicate good numerical convergence of the considered runs.

Modes with higher frequency, (5,5), (6,6), (7,7), (8,8),
unfortunately, appear often compromised during merger and ringdown.
The most common symptom is that these modes reach their maximum a few
$M$ earlier than expected, and do not exhibit a clear exponential
decay thereafter.  These symptoms are consistent with an
insufficiently fine radial grid, on which the short-wavelength
high-frequency merger waves would not be resolved well enough as they
propagate to the extraction spheres, and are thus unphysically damped
away.  Moreover, extrapolation to infinite extraction radius appears
to magnify non-physical features in these high-frequency modes, in
about half of the simulations considered. Gravitational wave extrapolation is most
important for the early inspiral, where the wavelength is
long~\cite{NRPNCaltech07,Boyle:2009vi} and is less important for the
merger portion considered here. To mitigate impact on the high-$m$
modes --(5,5) and above--, we therefore decided in the present study
to utilize the gravitational waveforms extracted at the largest
available extraction radius.

The impact on our analysis of imperfections in the underlying numerical data
can be judged in two ways: First, Figs.~\ref{fig:non_spinning_plots},
\ref{fig:aligned_plots_PN_time_series}, and
\ref{fig:aligned_plots_PN_mass_ratio_series} show error bars for each
data-point, obtained from the difference in value from numerical simulations of
different numerical resolution.  These error bars are generally small compared
to the physical effects being analysed.  Second, our analysis should produce
results that are slowly and smoothly varying with change of the underlying
binary black hole parameters like mass-ratio or black hole spin. In particular,
simulations at nearby parameter points should yield similar answers, and indeed
they do, even if the simulations come from very different epochs.  The results
obtained here (e.g. in Fig.~\ref{fig:non_spinning_plots}) vary smoothly with
parameters, with the scatter being consistent with the error bars.  As such, we
believe the underlying numerical data to be reliable for our purposes, except
perhaps, for the analyses of $(5,5)$ through $(8,8)$ in the regime after the
$(2,2)$ mode reached peak amplitude.

\section{Leading order post-Newtonian approximations of gravitational wave mode
amplitudes}\label{sec:PN}

The `plus' and `cross' gravitational wave polarizations can be decomposed in terms of spherical
harmonics as shown in Eq.~\eqref{eq:polarisations_and_modes}. The complete PN
expressions for various spherical harmonic modes $h_{\ell m}$, given the
currently available accuracies of the multipole moments, are reported in
Refs.~\cite{BFIS08} and \cite{ABFO08,Mishra:2016whh}, for nonspinning binaries
moving in circular orbits and for systems whose spins are aligned or
anti-aligned with respect to the orbital angular momentum, respectively.

As we are going to crucially exploit the leading order dependencies of these
modes, we list them below for convenience. Note that in these expressions $v$
is the PN velocity parameter and $\eta$ the symmetric mass ratio
defined as $\eta=\frac{m_1 m_2}{M^2}$, with binary component masses $m_1,m_2$
and total mass $M$.

The structure of the various modes in PN theory reads as~\cite{BFIS08}
\begin{subequations} \label{eq:PN_functions}
\begin{align}
h_{22}&=\mathcal{C}_{22}\,v^{2}\,\E^{-\I 2\psi}\,\eta\,\left (1+O(v^{2})\right ) \Tc \\
h_{21}&=\mathcal{C}_{21}\,v^{2}\,\E^{-\I \,\psi}\,\eta \,\left (\delta\,v+ O(v^{2})\right ) \Tc \\
h_{33}&=\mathcal{C}_{33}\,v^{2}\,\E^{-\I 3\psi}\,\eta \,\left (\delta\,v+O(v^{3})\right ) \Tc \\
h_{32}&=\mathcal{C}_{32}\,v^{2}\,\E^{-\I 2\psi}\,\eta \,\left ((1-3\,\eta)\,v^{2}+O(v^{3})\right ) \Tc \\
h_{31}&=\mathcal{C}_{31}\,v^{2}\,\E^{-\I \,\psi}\,\eta \,\left (\delta\,v+O(v^{3})\right ) \Tc \\
h_{44}&=\mathcal{C}_{44}\,v^{2}\,\E^{-\I 4\psi}\,\eta\,\left ((1-3\,\eta)\,v^{2}+O(v^{4})\right ) \Tc \\
h_{43}&=\mathcal{C}_{43}\,v^{2}\,\E^{-\I 3\psi}\,\eta\,\left (\delta\,(1-2\,\eta)\,v^{3}+O(v^{4})\right ) \Tc \\
h_{42}&=\mathcal{C}_{42}\,v^{2}\,\E^{-\I 2\psi}\,\eta\,\left ((1-3\,\eta)\,v^{2}+O(v^{4})\right ) \Tc  \\
h_{41}&=\mathcal{C}_{41}\,v^{2}\,\E^{-\I \,\psi}\,\eta\,\left (\delta\,(1-2\,\eta)\,v^{3}+O(v^{4})\right ) \Tc  \\
h_{55}&=\mathcal{C}_{55}\,v^{2}\,\E^{-\I 5\psi}\,\eta \,\left (\delta\,(1-2\,\eta)\,v^{3}+O(v^{5})\right ) \Tc \\
h_{66}&= \mathcal{C}_{66}\,v^{2}\,\E^{-\I 6\psi}\,\eta \,\left ((1-5\,\eta+5\eta^{2})\,v^{4}+O(v^{6})\right ) \Tc   \\
h_{77}&= \mathcal{C}_{77}\,v^{2}\,\E^{-\I 7\psi}\,\eta\,\left (\delta \,( 1-4\,\eta + 3 \eta^{2})\,v^{5}+O(v^{7})\right ) \Tc  \\
h_{88}&= \mathcal{C}_{88}\,v^{2}\,\E^{-\I 8\psi}\,\eta \,\left (( 1-7\eta+14\eta^{2}-7\eta^{3})\,v^{6}+O(v^{7})\right ) \Tc 
\end{align}
\end{subequations}
where $\mathcal{C}_{\ell m}$ are complex constants, $v$ is the PN velocity parameter which
captures the time dependency of the wave modes,
$\psi$ is the PN phase variable, and
$\delta=\frac{m_1-m_2}{m_1+m_2}$ is a mass asymmetry parameter which can be rewritten as 
$\delta=\sqrt{1-4\,\eta}$ for $m_1>m_2$. It vanishes for equal mass binaries.

Based on the structure of the expressions in Eqs.~\eqref{eq:PN_functions} we 
introduce the \emph{leading order PN approximations} which capture the leading order 
$\eta$ and spin dependencies for fixed $v$---i.e. at a fixed 
time---and thus allow us to examine the numerical relativity waveforms for \emph{PN signature} or rather
deviations from it. Our goal is somewhat diffent from the usual approach in the literature 
as we are aiming to study the behavior of the mode amplitudes in terms of the intrinsic 
parameters of the binary system and not as a function of time.
\subsection{Nonspinning binaries}
In order to gain insight into the behavior of the amplitudes of the nonspinning modes, we choose the following 
fitting functions $A_{\ell m}=|h_{\ell m}|$ which capture the leading order dependencies 
of the PN expressions~\eqref{eq:PN_functions} on the mass ratio parameters $\eta$ and $\delta$:
\begin{subequations} \label{eq:fit_functions}
\begin{align}
A_{22} &= \alpha_{22} \,\eta \Tc \\
\hat{A}_{21} &= \alpha_{21} \,\delta \Tc \\
\hat{A}_{33} &= \alpha_{33} \,\delta \Tc\\
\hat{A}_{32} &= \alpha_{32} \,\left (1-3\,\eta\right )\Tc \\
\hat{A}_{31} &= \alpha_{31} \,\delta \Tc \\
\hat{A}_{44} &= \alpha_{44} \,\left (1-3\,\eta\right )\Tc \\
\hat{A}_{43} &= \alpha_{43} \,\delta\, \left (1-2\,\eta\right )\Tc  \\
\hat{A}_{42} &= \alpha_{42} \,\left (1-3\,\eta\right )\Tc  \\
\hat{A}_{41} &= \alpha_{41} \,\delta\, \left (1-2\,\eta\right )\Tc \\
\hat{A}_{55} &= \alpha_{55} \,\delta\, \left (1-2\,\eta\right )\Tc   \\
\hat{A}_{66} &= \alpha_{66} \,\left (1-5\,\eta+5\eta^{2}\right )\Tc  \\
\hat{A}_{77} &= \alpha_{77} \,\delta\,\left ( 1-4\,\eta + 3 \eta^{2}\right ) \Tc \\
\hat{A}_{88} &= \alpha_{88} \,\left ( 1-7\eta+14\eta^{2}-7\eta^{3}\right )\Tc  
\end{align}
\end{subequations}
where $\alpha_{\ell m}$ are the scaling factors that we fit for. The hatted
amplitudes $\hat{A}_{\ell m}={A_{\ell m}}/{A_{22}}$ have been normalized
with respect to the (2,2) mode to cancel the overall $\eta$-factor present
in every mode.
\subsection{Aligned spin binaries} \label{sec:aligned_PN}
In PN theory, spin effects are sub-dominant and are not present at leading order for any 
mode \cite{Mishra:2016whh}. Current-multipole modes which obey $\ell+m=\text{odd}$ contain 
spin-dependent terms at 0.5 PN order above the leading term and thus are more likely to 
exhibit spin effects \cite{Kamaretsos2012b}. We focus on the four current-multipole modes with 
$\ell\leq 4$, $(2,1)$, $(3,2)$, $(4,3)$, and $(4,1)$.
The PN expression for the (2,1) mode to the next-to-leading order in $v$ is 
given by
\begin{equation}
h_{21}=C(v,\psi)\,\eta \,\left (\delta\, v - \dfrac{3}{2}\left ( \bm{\chi_{a}}+\delta\,\bm{\chi_{s}}\right )\cdot\bm{\hat{L}_{N}}\,v^{2}\right ) + O(v^{3}) \Tc \label{eq:PN_spin_21}\\ 
\end{equation}
where $C$ is a function of the orbital velocity $v$ and the PN phase
variable $\psi$, $\bm{L_{N}}$ is the orbital angular momentum, and
\begin{align}
  \bm{\chi_{s}}&=\frac{1}{2}(\bm{\chi_{1}}+\bm{\chi_{2}})\\
  \bm{\chi_{a}}&=\frac{1}{2}(\bm{\chi_{1}}-\bm{\chi_{2}})
\end{align}
denote, respectively,  the symmetric and antisymmetric spin combinations of the initial black hole 
spins $\bm{\chi_{1}}$ and $\bm{\chi_{2}}$. Since we assume that spins and the orbital 
angular momentum are aligned, we can write instead
\begin{equation}
h_{21}=C(v,\psi)\,\eta \,\left (\delta\, v - \dfrac{3}{2}\,\chi^{\eff}_{21}\,v^{2}\right ) + O(v^{3}) \Tc \\ 
\end{equation}
where $\chi^{\eff}_{21}=\chi_{a}+\delta\,\chi_{s}$,
with $\chi_{a,s}=\bm{\chi}_{a,s}\cdot\bm{\hat L_N}$ being the projection of the 
symmetric/antisymmetric spin vectors along the orbital angular momentum.
This form motivates the fitting ansatz, with non-spinning
$\hat{A}_{21}^{\text{ns}}=\delta$ from \eqref{eq:fit_functions}, which reads as
\begin{align}
\hat{A}_{21}=\gamma_{21} \, \hat{A}_{21}^{\text{ns}}+ \beta_{21} \,
\chi^{\eff}_{21} \Td 
\end{align}
This can be generalized to arbitrary $\ell m$ as
\begin{align}
\hat{A}_{\ell m}=\gamma_{\ell m} \,\hat{A}^{\text{ns}}_{\ell m}(\eta) + \beta_{\ell m} \, \chi^{\eff}_{\ell m}(\eta,\chi_{1},\chi_{2}) \Tc \label{eq:spin_fits}
\end{align}
with different effective spin parameters for
different modes defined by the linear combination of the spin
parameters in the PN expressions for those modes.
The functional forms of the effective spin parameters for the different
modes are motivated by Eqs.~(12) of  Ref.~\cite{ABFO08} and are given
by
\begin{subequations}
 \label{eq:effective_PN_spins}
\begin{align}
\chi^{\eff}_{21} &= \chi_{a}+\delta\,\chi_{s}\Tc\\
\chi^{\eff}_{32} & = \eta \,\chi_{s}\Tc\\
\chi^{\eff}_{43} = \chi^{\eff}_{41} & = \eta \left ( \chi_{a}-\delta\,\chi_{s}\right ) \Td
\end{align}
\end{subequations}
Equation~\eqref{eq:spin_fits} has two fit parameters $\beta_{\ell m},\gamma_{\ell m}$
whereas Eqs. \eqref{eq:fit_functions} only need one. The additional
parameter is aimed to capture the extra degrees of freedom due to
spins and account for the fact that the nonspinning and spinning effects
enter at different PN orders.

\section{Post-Newtonian signature in numerical relativity waveform amplitudes}
\label{sec:PNSig}
\begin{figure*}
    \includegraphics[width=\textwidth]{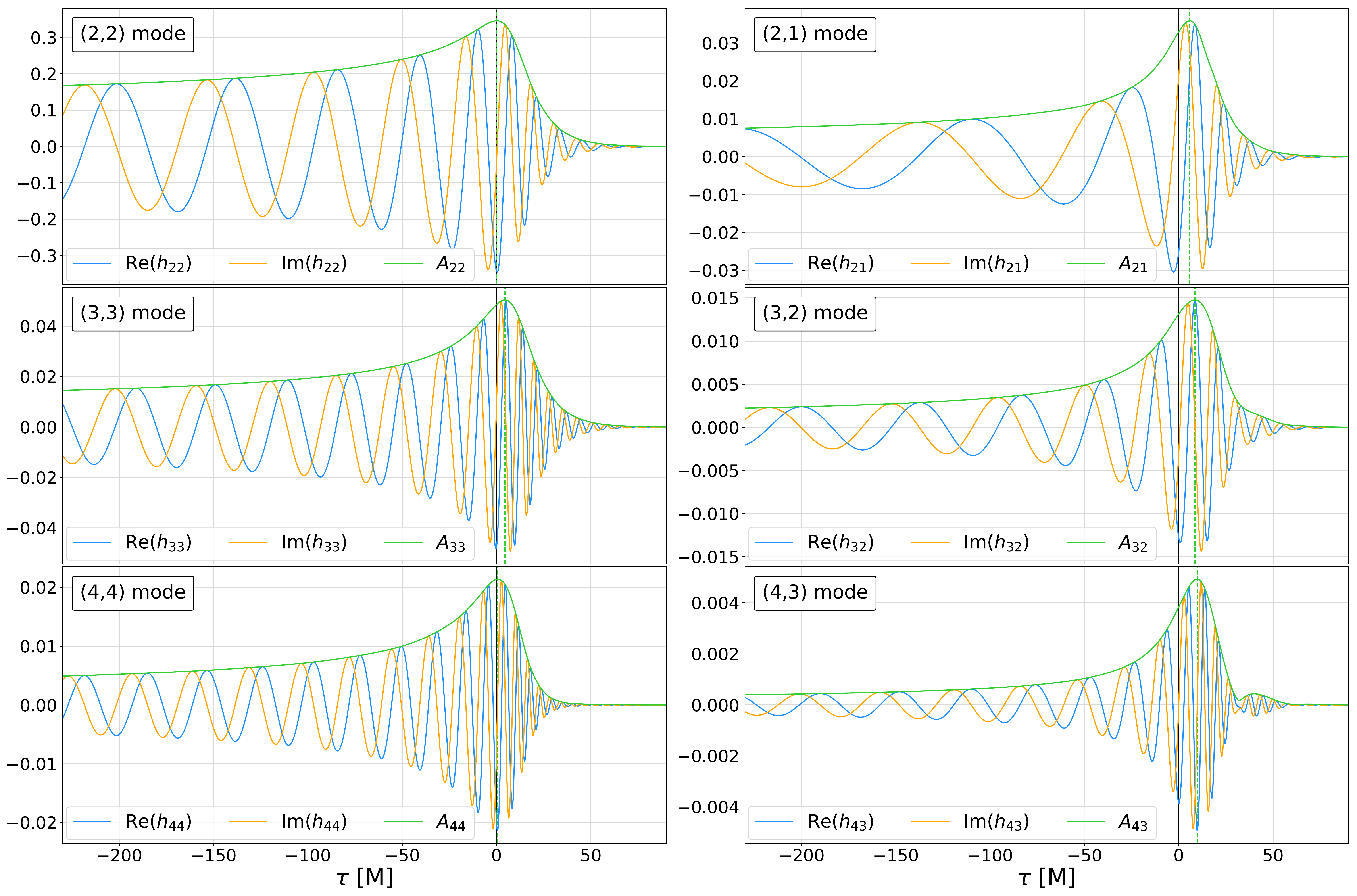}
    \caption{The SXS waveform catalog provides access to many compact bianry simulations. As a representation of the catalog, we show the numerical data of six wave modes $h_{\ell m}$ from the non-spinning run SXS:BBH:0169, with mass ratio $q=2$. The figure plots the real and imaginary parts of the wave modes and their amplitude $A_{\ell m}=\sqrt{(\operatorname{Re}(h_{\ell m}))^{2}+(\operatorname{Im}(h_{\ell m}))^{2}}$ for the (2,2), (2,1), (3,3), (3,2), (4,4), and (4,3) modes as a function of $\tau=t-t_{\text{max}}^{22}$. The position of the peak amplitude of the (2,2) mode is highlighted with a solid, black line, while the dashed, green line corresponds to the maximum of each mode.}
    \label{fig:sxs_wavemodes_22_21_33_32_44_43}
\end{figure*}

\subsection{Numerical wave mode amplitudes}
From each of the numerical relativity simulations summarized in Sec.~\ref{sec:SXS} and detailed in Appendix~\ref{app:SXS_simulations}, we extract the time series of real and imaginary components of the spherical harmonic wave modes $h_{\ell m}$ (see Eq. \eqref{eq:polarisations_and_modes}) for the 13 modes with $(\ell,m)\in\{$(2,2), (2,1), (3,3), (3,2), (3,1), (4,4), (4,3), (4,2), (4,1), (5,5), (6,6), (7,7), (8,8)$\}$. The remaining numerical modes with $\ell \geq 5, m\neq \ell$ are excluded from this study due to their relatively small amplitudes and large numerical errors. Fig.~\ref{fig:sxs_wavemodes_22_21_33_32_44_43} shows the real and imaginary components of six of the 13 wave modes together with their amplitude for an example SXS run: SXS:BBH:0169, mass ratio $q=2$, and non-spinning.

Our goal is to examine the behavior of various spherical harmonic
modes for PN signature and deviations from it around the time of
merger. We concentrate this study on the evolution of the real amplitude
\begin{equation} \label{eq:wavemode_amplitude}
  A_{\ell m}=\sqrt{\operatorname{Re}(h_{\ell m})^{2}+\operatorname{Im}(h_{\ell m})^{2}},
\end{equation}
of the spherical harmonic wave modes $h_{\ell m}$, where both $A_{\ell m}$ and $h_{\ell m}$ are functions of time $\tau$, as well as binary parameters $M,\eta,\vec{\chi}_{1},\vec{\chi}_{2}$.
The time variable $\tau=t-t_{\text{max}}^{22}$ has been shifted such that the
peak amplitude of the (2,2) mode is located at $\tau=0$. In the case of
non-spinning binary black holes with quasi-circular orbits the parameter tuple
$\theta=(M,\eta,\vec{\chi}_{1},\vec{\chi}_{2})$ reduces to the two mass
parameters, the total mass $M$, and the symmetric mass ratio $\eta$. If the black holes
are aligned spinning, $\theta=(M,\eta,\chi_1,\chi_2)$ also contains the spin
magnitudes.
\subsection{Results for non-spinning binaries}
\begin{figure*}
    \includegraphics[width=0.95\textwidth]{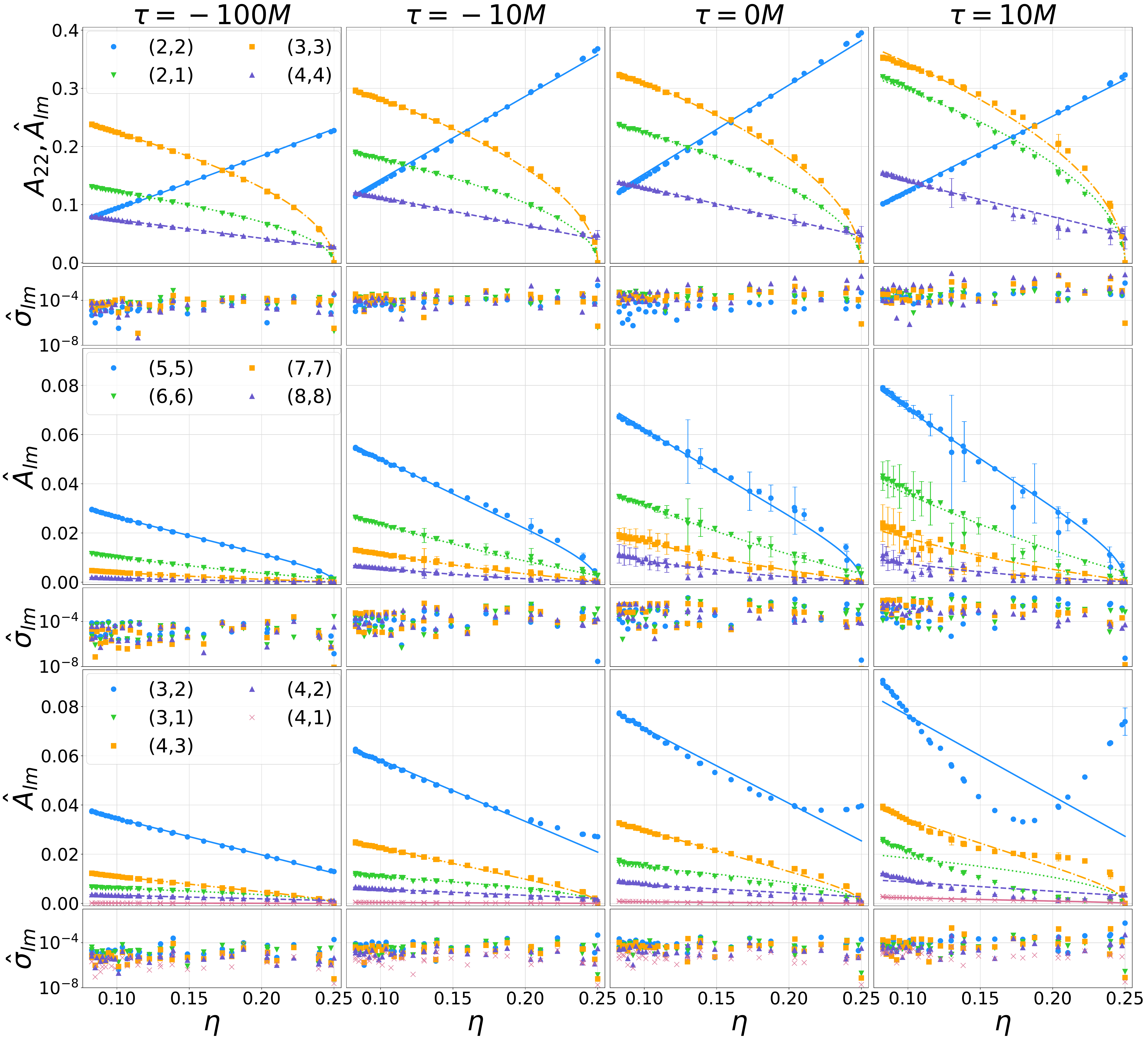}
    \caption{\emph{Non-spinning simulations:} The PN inspired amplitude fits for 13 spherical harmonic modes (\emph{lines}) with spin weight $-2$ are plotted together with the numerical data (\emph{points}) of simulations of the coalescence of two non-spinning black holes at four different times $\tau=t-t_{\text{max}}^{22}$. The data has been spread over 24 subplots with columns representing different times $\tau/M=-100, -10, 0, 10$ and rows grouping the modes by strength, fit agreement, and numerical errors.  The 12 large plots present the relative amplitudes $\hat{A}_{\ell m}\equiv \frac{A_{\ell m}}{A_{22}}$ against the symmetric mass $\eta$, with $A_{22}$ as an exception, while the accompanying smaller plots show the numerical errors of the simulations which are also visible as error bars in the main plots. The plots show that the four dominant and the $\ell=m$ modes maintain a PN-like signature throughout the studied time range, while the $\ell\neq m$ modes start to deviate from this PN-like behavior and thus capture the deviations from the PN description most efficiently.}
    \label{fig:non_spinning_plots}
\end{figure*}
\begin{figure*}
    \includegraphics[width=0.95\textwidth]{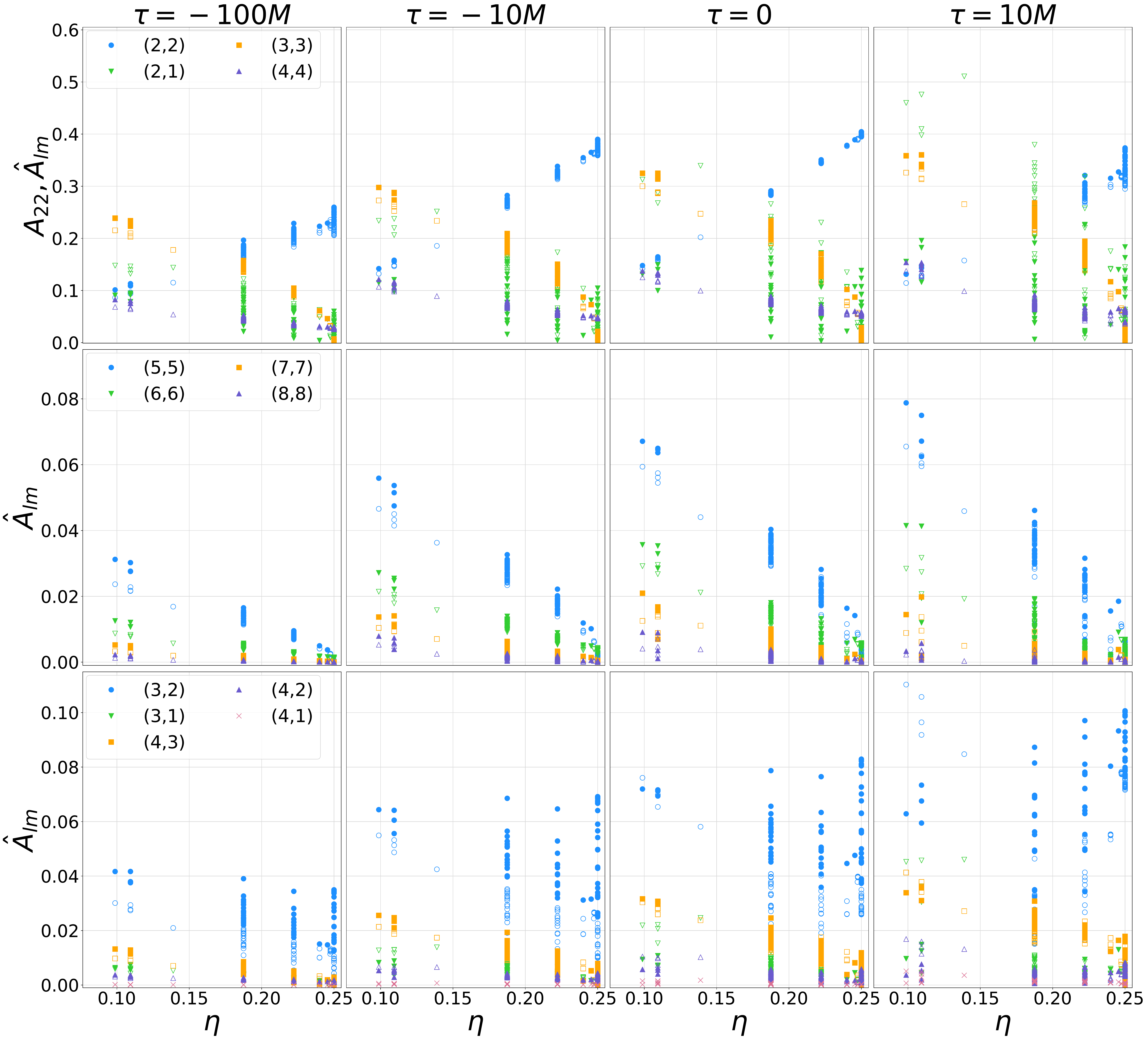}
    \caption{\emph{Aligned spin simulations:} The relative amplitudes $\hat{A}_{\ell m}\equiv \frac{A_{\ell m}}{A_{22}}$ of 13 spherical harmonic modes with spin weight $-2$ are plotted against the symmetric mass $\eta$, with $A_{22}$ as an exception. The data was taken from 121 numerical simulations of the coalescence of two black holes whose spins is aligned with respect to the orbital angular momentum of the binary. The amplitudes are shown at four different times, $\frac{\tau}{M}=-100,-10,0,10$, $\tau=t-t_{\text{max}}^{22}$. The data is presented by 12 subplots dividing the modes in three groups by strength. The vertical spread at a fixed symmetric mass ratio indicates that the spin information cannot be captured in an one dimensional plot over the symmetric mass ratio (compare against Fig.~\ref{fig:non_spinning_plots}). The effective spin is the standard symmetric, mass weighted version: $\chi_{\eff}=\frac{m_1 \chi_1 + m_2 \chi_2}{M}$ (\emph{full symbol}: $\chi_{\eff}\geq 0$, \emph{empty symbol}: $\chi_{\eff}< 0$).\\The amplitudes of the (2,1) and (3,2) modes show especially large variations at a given value of $\eta$, hence pointing towards their strong dependence on the spin properties of the system. From PN theory we would also expect any other mode with $\ell+m=\text{odd}$ to have a strong spin dependence.}
    \label{fig:aligned_plots_PN_all_modes_spread}
\end{figure*}
Figure~\ref{fig:non_spinning_plots} contains the condensed results of
our study of PN signature in the waveform amplitudes of initially
non-spinning binary black holes around the time of merger.
The individual symbols in these plots are the amplitudes of the various gravitational wave modes extracted from the SXS simulations.
Each symbol carries an error bar, often too small to be visible, that is derived as the difference in the extracted amplitude for at least two different numerical resolutions.
The solid lines in Fig.~\ref{fig:non_spinning_plots} represent our leading order
PN approximations in Eq.~\eqref{eq:fit_functions} fitted to the numerical data for various wave mode 
amplitudes as a function of the symmetric mass ratio. The temporal evolution is presented via snapshots 
at four different times ${\tau}/{M}=-100,-10,0,10$, corresponding to the columns in Fig.~\ref{fig:non_spinning_plots}.
The three rows group the different modes by numerical strength. The first row contains the data and fits for the four strongest 
modes (2,2), (2,1), (3,3), and (4,4), the second row shows the remaining $\ell=m$ modes, and the sub-dominant modes with 
$\ell = 3, 4 $ are bundled in the last row. The larger subfigures plot relative amplitudes 
$\hat{A}_{\ell m}(\tau)={A_{\ell m}(\tau)}/{A_{22}(\tau)}$, $\ell m\neq 22$, with only $A_{22}$ being shown as an absolute
amplitude. The error bars correspond to twice the numerical errors $\sigma_{\ell m}$ shown in the smaller subplots.

\indent The amplitude of the (2,2) mode behaves as expected and increases
towards its maximum at $\tau=0$. Due to the suppression of the two next
strongest modes, (3,3) and (2,1), for equal mass binaries, the (2,2)
mode stays most significant in the realm between $q=1$ to $q=2$ which is
where all detections by the LIGO Virgo Collaboration were
made~\cite{O1BBH,GW170104,GW170814,GW170608}. The situation for the other modes paints a more interesting picture for low symmetric mass ratios $\eta<0.15$ where their amplitudes increase more quickly relative to the (2,2) mode. This tendency shows the importance of the inclusion of higher modes for medium to extreme mass ratio binary coalescences.
\\
\indent The first column in Fig.~\ref{fig:non_spinning_plots} shows the 
comparison of the leading order PN approximations to the numerical data at time 
$\tau=-100 M$. The approximations work beautifully and confirm the expectation that PN 
theory describes the functional dependence of the gravitational wave amplitudes on $\eta$ very well 
during the inspiral. The level of agreement between the data and the fits is quantified 
by the correlation coefficients\footnote{The correlation coefficient $C$ between the data vector $\bm d$ and
the appropriate fit vector $\bm f$, with averages $\bar{d},\bar{f}$, is defined as
$C =\frac{(\bm{f}-\bar{f})\cdot(\bm{d}-\bar{d})}{\sqrt{(\bm{f}-\bar{f})^2(\bm{d}-\bar{d})^2}}$.}
in Table \ref{tab:corr_coeff}. The situation stays very similar close to merger at $\tau=-10M$, even
though the amplitude of the (3,2) mode is starting to show deviations
from the PN inspired fitting. The fits for the remaining modes capture the data extremely 
well despite the common belief that PN theory should fail in this regime due to the 
increase in the orbital velocity parameter $v$.
\\
\begin{table}
    \centering
    \begin{tabular}{|c|c|c|c|c|}
    \hline 
    Mode & $\tau=-100M$ & $\tau=-10M$ & $\tau=0M$ & $\tau=10M$ \\ 
    \hline \hline
    (2,2) & 0.999998 & 0.999779 & 0.999703 & 0.999798 \\
    \hline
    (2,1) & 0.999579 & 0.999854 & 0.999838 & 0.998717 \\
    \hline
    (3,3) & 0.999912 & 0.999787 & 0.999301 & 0.997751 \\
    \hline
    (4,4) & 0.999667 & 0.999299 & 0.998743 & 0.991130 \\
    \hline \hline
    (5,5) & 0.999842 & 0.999204 & 0.997655 & 0.993545 \\
    \hline
    (6,6) & 0.999722 & 0.998421 & 0.994847 & 0.976633 \\
    \hline
    (7,7) & 0.999698 & 0.996487 & 0.982980 & 0.939223 \\
    \hline
    (8,8) & 0.999491 & 0.994833 & 0.965470 & 0.848470 \\
    \hline \hline
    (3,2) & 0.999376 & 0.996502 & 0.972295 & 0.585903 \\
    \hline
    (3,1) & 0.997824 & 0.993977 & 0.981189 & 0.908449 \\
    \hline
    (4,3) & 0.999719 & 0.999168 & 0.998515 & 0.974125 \\
    \hline
    (4,2) & 0.998844 & 0.997680 & 0.995446 & 0.908354 \\
    \hline
    (4,1) & 0.984923 & 0.948784 & 0.863388 & 0.976998 \\
    \hline
    \end{tabular}
    \caption{Correlation coefficients of the non-spinning fits to the SXS data for spherical harmonic modes (2,2), (2,1), (3,3), (3,2), (3,1), (4,4), (4,3), (4,2), (4,1), (5,5), (6,6), (7,7), and (8,8) at $\frac{\tau}{M}=-100,-10,0,10$.}
    \label{tab:corr_coeff}
\end{table}
\indent The picture becomes truly exciting at $\tau=0$, after a common horizon has already
formed. The amplitudes of the four dominant
modes are fitted exceptionally well by the leading order PN approximations,
hence giving us an insight into how little these amplitudes are affected
by the dynamics during the coalescence of the binary system. The
PN-inspired fits to the $\ell=m$ modes in the second row of Fig.~\ref{fig:non_spinning_plots} 
are still remarkably well captured,
especially for $\ell=5,6$. The amplitudes of the modes with $\ell=3,4$,
$m<\ell$ exhibit a different behavior: their numerical amplitudes deviate
strongly from the PN-inspired fitting and thus indicating that the merger process affects 
the dynamics of these mode amplitudes more than the four dominant modes or the modes with $\ell=m$.
\\
\indent Finally, the last column of Fig.~\ref{fig:non_spinning_plots} contains
the data and fits during the early ringdown at $\tau=10M$. The deviations from
the leading order PN approximations have increased, compared to time $\tau=0$
which is reflected in the correlation coefficients in Table
\ref{tab:corr_coeff}. The amplitude data for $\tau=10M$ in the second row of
Fig.~\ref{fig:non_spinning_plots} appears to be well captured by the leading
order PN approximations, but it exhibits large numerical errors that make a
quantitative evaluation of the approximations impractical, see Table
\ref{tab:corr_coeff}. The dominant modes show a very intriguing outcome. Their
amplitudes seem to maintain the PN signature from earlier times fairly well.
This reproduces the earlier findings \citep{Kamaretsos2012a} that found
$\eta$-dependences in these amplitudes during the ringdown. Our analysis goes
beyond that and shows that this dependence is still mostly of PN signature
$10M$ after the merger.  \\
\indent In summary, we can say that the four dominant modes with large
amplitudes and the wave modes with $\ell=m$ maintain the PN signature of the
inspiral phase exceptionally well in their amplitudes, from the inspiral throughout the merger into
the ringdown, while the spherical harmonic wave modes with $\ell \neq m$
deviate from this PN-like behavior as the evolution of the binary approaches
the merger. Thus, these present interesting candidates for binary black hole
merger studies and strong field tests of general relativity, with the (3,2) mode being especially
intriguing as it is the strongest of these modes and hence the most significant
for future detections.
\subsection{Results for aligned spins}
\begin{figure*}
    \includegraphics[width=0.95\textwidth]{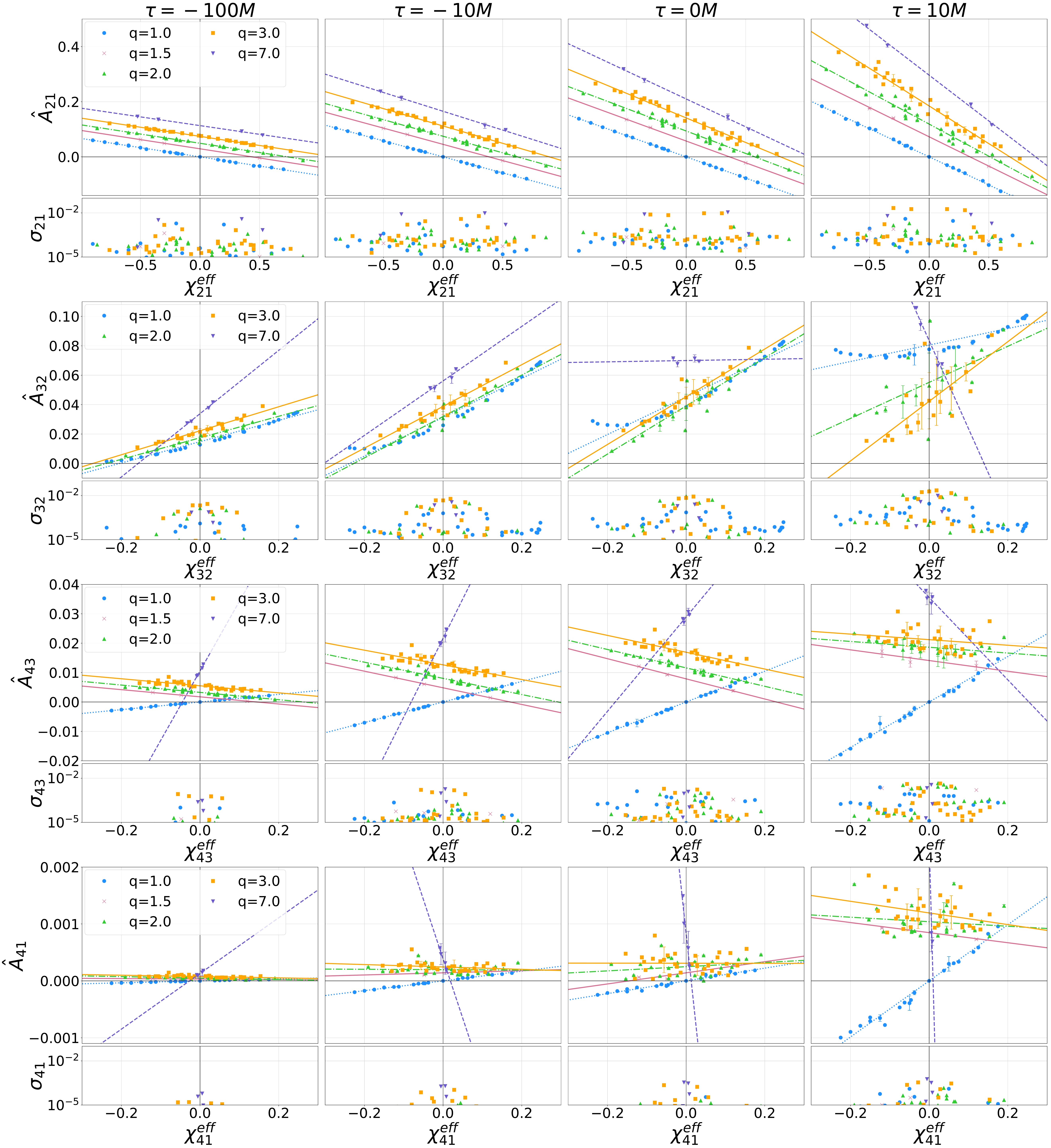}
    \caption{The relative amplitudes $\hat{A}_{lm}=A_{lm}/A_{22}$ of
      the (2,1), (3,2), (4,3), and (4,1) modes are presented against
      their respective effective spin combinations
      $\chi^{\eff}_{lm}$. The columns represent four times
      $\tau/M=-100,-10,0,10$, and the $\chi_{lm}^{\rm eff}$ are
      defined in Eqs.~(\ref{eq:effective_PN_spins}). The plotted modes with
      $l+m=\text{odd}$, have low order effective spin contributions in PN
      theory due to their current-multipole nature. The PN inspired fits
      (\emph{lines}) in Eq. \eqref{eq:spin_fits} are linear in
      these effective spin combinations, thus three data points
      (\emph{SXS data}) give an indication as to whether the
      amplitudes behave in a PN-like way. The figure consists of four major
      rows, each dedicated to one of the wave modes, and four columns,
      capturing the times, with subfigures showing the relative
      amplitudes and in a smaller window the error of the numerical
      data. The restriction to the four mass ratios $q=1,1.5,2,3,7$ is
      due to the requirements for numerical error estimation and
      having three simulations with different effective spins
      $\chi_{lm}^{\eff}$ per mass ratio for the linear fits. For the
      available set of simulations with $q=1.5$, symmetry reduces, in
      the case of the (3,2) mode, the three different sets of $\chi_1$
      and $\chi_2$ to two.}
    \label{fig:aligned_plots_PN_time_series}
\end{figure*}

Let us now discuss the aligned-spin simulations.  We extract amplitudes $A_{22}$ and $\hat A_{\ell m}$ as above for all simulations with aligned spins.
Fig.~\ref{fig:aligned_plots_PN_all_modes_spread} plots the
amplitudes of all 13 modes for each aligned spinning SXS simulations
in a distinct data point.
For a given value of $\eta$, there are generally multiple
simulations with different spins; these simulations lead to different amplitudes, resulting in the vertical scatter of data-points at each $\eta$.
Most modes show a small amount of vertical scatter
whereas the (2,1) and (3,2) mode exhibit significant variation amongst
the different simulations at a given symmetric mass ratio.  This
large amount of spread in the
amplitudes of the aforementioned modes is a consequence of a strong
dependence on the omitted two parameters, the spin magnitudes $\chi_1$
and $\chi_2$ of the component black holes.

Not only do the (2,1) and (3,2) modes show the most pronounced
scatter in Fig.~\ref{fig:aligned_plots_PN_all_modes_spread}, but
they are also among the current-multipole modes with
$\ell+m=\text{odd}$, which exhibit spin effects at low PN order (cf. Sec.~\ref{sec:aligned_PN}).  We will therefore now
investigate the spin dependence of the relevant four modes (2,1),
(3,2), (4,3), and (4,1) in more detail. The results and
corresponding correlation coefficients for the amplitude data and the
PN-inspired fits for the (2,1), (3,2), (4,3), and (4,1) modes are presented
in Fig.~\ref{fig:aligned_plots_PN_time_series} and Table
\ref{tab:corr_coeff_as}, respectively. While Fig.~\ref{fig:aligned_plots_PN_time_series} spreads the temporal evolution
of the mode amplitudes for given mass ratios $q$ in columns, we give
another representation of the same information in Fig.~\ref{fig:aligned_plots_PN_mass_ratio_series} in Appendix
\ref{app:alternative_mass_ratio_series}, where the roles of mass ratio
$q$ and time $\tau$ are flipped. This gives a better intuition of how
each mode amplitude evolves for fixed mass ratio throughout the
merger.  \\ \indent The (2,1) modes appears to behave similarly to the
non-spinning case and maintain the PN signature of the inspiral
throughout the merger into the early ringdown. The leading order PN
approximations capture the numerical relativity data exceptionally well for mass ratios
$q=1,1.5,7$. Mass ratios $q=2,3$ show a larger scatter around the
linear fit line which appears to be a result of the larger variation
of initial spin combinations of the SXS waveforms for lower mass
ratios, see Table \ref{tab:simulations_as} (in particular,
the SXS simulations explore variations of anti-symmetric spin $\chi_1-\chi_2$ much more exhaustively for $q=1,2,3$). This scatter is not
visible in the data for mass ratio $q=1$ due to the suppression of
asymmetries for equal mass binaries.  \\ \indent The results are
similar for the (4,3) and (4,1) modes. Mass ratio $q=1$ is again
extremely well captured by the leading order PN approximations. While mass
ratios $q=1.5,7$ appear to be fitted well in Fig.~\ref{fig:aligned_plots_PN_time_series}, the correlation coefficients
show that the PN inspired fit performs less admirably than for the (2,1)
mode. The scatter around the approximation lines for $q=2,3$ is also
much more prevalent, especially at times $\tau/M=0,10$.
\\ \indent Our analysis of the (3,2) mode does not include mass ratio
$q=1.5$: Two of the three distinct initial spin combinations reduce to
the same effective spin $\chi^{\eff}_{32} = \eta \,\chi_{s}$ due to
its symmetry and thus make a linear fit to two points a moot
exercise. The analysis of the (3,2) mode shows that it takes a special
role amongst the four studied modes when the system includes
spins. The (3,2) mode amplitude is the only to show major deviations
from the PN signature for equal mass binary waveforms, $q=1$, and even
during the inspiral at $\tau=-100M$. The picture for mass ratios
$q=2,3,7$ is the same as what we observed for the (4,3) and (4,1)
modes.
\\
\indent In summary, the (2,1) mode seems to do as well as
we saw from the non-spinning scenario, while the three weaker modes
exhibit various different effects. All modes, even (2,1), showed some
level of scatter for mass ratios $q=2,3$ that probably is a result of
variations in the initial spin data, which did not show up for equal
mass binaries for the three modes with odd $m$, (2,1), (4,3), and
(4,1). Hence, it appears that the addition of aligned spin has a
stabilizing effect against deviations from a PN signature for these
modes, if the mass ratio is $q=1$. The (3,2) mode takes a distinct
role as its amplitude for mass ratio $q=1$ shows deviations from the
leading order PN approximation already during the inspiral. Hence, it again stands
out as the mode of interest in strong field tests of general relativity.
\section{Implications of the results for modelling waveforms from binary black holes}\label{sec:Implications}
In this Section we will discuss the behavior of the various modes as
a function of time and the system's mass ratio $\eta$ and mode-dependent `effective spin'
(which is different for different modes).

\paragraph{Time evolution} Our study has shown that for most part of the
adiabatic evolution when $\dot \omega/\omega^2 \ll 1,$ where $\omega$ is the
orbital frequency, the mass-ratio dependence of the amplitudes of the various
spherical harmonic modes are as predicted by PN theory to lowest
order. The overall multipole structure is set in when the two black holes are
well-separated; it is difficult to deform this multipolar structure because
spacetime has a large bulk modulus. Remarkably, most $\ell=m$ mode amplitudes
continue to agree with the PN prediction well after the common
horizon has formed. This includes the dominant $\ell=m=2,3,4$ modes as well as
the weaker $\ell=m>4$ modes (cf. Fig.~\ref{fig:non_spinning_plots}, first and
second row). The strong field dynamics does affect the $\ell\ne m$ modes
(except the strong  $(2,1)$ mode), especially when the two bodies get closer
together as evidenced by the change in the {\em weaker} $\ell \ne m$ modes (cf.
Fig.~\ref{fig:non_spinning_plots}, last row). Our analysis shows that as we get
close to the merger phase, say $\tau \sim -10 M$ (this is about when the common
apparent horizon forms) the moments begin to deform from their PN
behavior.
\begin{table} \centering
    \begin{tabular}{|c|c|c|c|c|c|} \hline Mode & $q$ & $\tau=-100M$ &
    $\tau=-10M$ & $\tau=0M$ & $\tau=10M$ \\ \hline \hline
    \multirow{5}{*}{(2,1)} & 1 & 0.999325 & 0.999773 & 0.999988 & 0.999619 \\
    \cline{2-6} & 1.5 & 0.999980 & 0.999931 & 0.999807 & 0.999291 \\
    \cline{2-6} & 2 & 0.997846 & 0.995953 & 0.991763 & 0.984452 \\ \cline{2-6}
    & 3 & 0.997914 & 0.994403 & 0.991027 & 0.985538 \\ \cline{2-6} & 7 &
    0.999152 & 0.998667 & 0.999101 & 0.998829 \\ \hline \hline
    \multirow{5}{*}{(3,2)} & 1 & 0.990234 & 0.981084 & 0.968417 & 0.897599 \\
    \cline{2-6} & 2 & 0.980074 & 0.960144 & 0.914401 & 0.581929 \\ \cline{2-6}
    & 3 & 0.964765 & 0.952854 & 0.938971 & 0.784182 \\ \cline{2-6} & 7 &
    0.993511 & 0.953335 & 0.056356 & 0.980378 \\ \hline \hline
    \multirow{5}{*}{(4,3)} & 1 & 0.998952 & 0.999922 & 0.998798 & 0.995412 \\
    \cline{2-6} & 1.5 & 0.999844 & 0.999494 & 0.998659 & 0.874023 \\
    \cline{2-6} & 2 & 0.961789 & 0.958457 & 0.960024 & 0.472553 \\ \cline{2-6}
    & 3 & 0.844141 & 0.843969 & 0.837170 & 0.222383 \\ \cline{2-6} & 7 &
    0.987371 & 0.938732 & 0.759405 & 0.621659 \\ \hline \hline
    \multirow{5}{*}{(4,1)} & 1 & 0.991669 & 0.995539 & 0.985739 & 0.988915 \\
    \cline{2-6} & 1.5 & 0.608697 & 0.623753 & 0.916280 & 0.954856 \\
    \cline{2-6} & 2 & 0.684519 & 0.041862 & 0.220282 & 0.144261 \\ \cline{2-6}
    & 3 & 0.485186 & 0.233015 & 0.002791 & 0.334364 \\ \cline{2-6} & 7 &
    0.926836 & 0.992870 & 0.981711 & 0.982951 \\ \hline \end{tabular}
    \caption{Correlation coefficients of the aligned spinning fits to the SXS
    data for spherical harmonic modes (2,1), (3,2), (4,3), and (4,1) and mass
    ratios $q=1,1.5,2,3,7$ at $\frac{\tau}{M}=-100,-10,0,10$.}
    \label{tab:corr_coeff_as} \end{table}
\paragraph{$\eta$-dependence} Figures~\ref{fig:non_spinning_plots} and
\ref{fig:aligned_plots_PN_time_series} show the behavior of the mode amplitudes
as a fucntion of symmetric mass ratio $\eta$ at different epochs and as a
function of `effective spin' for different mass ratios $q$ and epochs,
respectively. For nonspinning systems, the $\ell=m$ modes are in pretty good
agreement with the leading order PN behavior as a function of
$\eta$ (see Eqs.\,\ref{eq:fit_functions}). This is true both at earlier times
$\tau\sim 100\,M$ when PN equations are expected to provide a good
description of the mode amplitudes, as well as at epochs when the
PN equations are believed not to be accurate. In fact, even at the
onset of merger at $\tau \simeq -10\,M$ and beyond $\tau =0$ when the black
hole begins to settle down (i.e. $\tau \sim 10\, M$) $\ell=m$ modes show little
departure from the PN behavior.

However, the weaker $\ell \ne m$ modes are altered significantly already at the onset
of the merger ($\tau \sim -10\,M$), especially for comparable mass binaries (i.e.
$\eta \simeq 1/4$). One exception to this rule is the $\ell=2,$ $m=1$ mode. This mode
is the strongest sub-dominant mode after $\ell=m=3$ (see Fig.\,\ref{fig:non_spinning_plots})
and is not easily modified by the strong field dynamics. The amplitude of the other
$\ell\ne m$ modes are at the level of $\lesssim 8\%$ (for highly asymmetric systems) of
the $(2,2)$ mode amplitude, while the $\ell=2,$ $m=1$ mode could be as large as 30\% of
the overall amplitude. 

\paragraph{Spin-dependence}
Figures~\ref{fig:aligned_plots_PN_time_series} and \ref{fig:aligned_plots_PN_mass_ratio_series}
present the mode amplitudes of the four $\ell + m = \text{odd}$ modes (2,1), (3,2),
(4,3), and (4,1) as a function of their respective effective spins (see Eq.
\ref{eq:effective_PN_spins}) for different mass ratios and epochs. Again, the
PN approximation does remarkably well at capturing the behavior of the
numerical data for the strong (2,1) mode, for all epochs and mass ratios. However,
the three weaker modes whose amplitudes show deviations from the PN
behavior for \emph{non-spinning} systems as early as $\tau \sim - 10\, M$, agree
with the PN approximation to some extent when the system includes
\emph{aligned spins}. Hence, it appears that the addition of spin to the system has
a stabilizing effect on the PN signature.

That being said, the situation is much more complicated than for non-spinning systems
as the quality of the agreement with the PN signature depends not only on
the mode and epoch, but also the mass ratio and the sample spread of initial black hole
spins. The agreement is good for mass ratios $q=1.5,7$ for which the data shows the
linear behavior in the respective effective spin combination for all modes.
Mass ratios $q=2,3$ were sampled with a much larger distribution in the initial
spins (cf. Table \ref{tab:simulations_as}), resulting in an envelope of data
points around the linear PN approximation in $\chi^{\eff}_{\ell
m}$. These envelopes widen during later epochs and for weaker modes. The found
spread in the data hints that the relationship between initial black hole spins
and mode amplitudes during the merger and ringdown cannot be captured in one
effective spin combination. Finally, the equal mass systems were sampled with a
similarly large spread in the intial spins, but do not show the envelope
characteristics of the higher mass ratios which is a result of the symmetry in
the system. The amplitudes of the three odd-$m$ modes (2,1), (4,3), and (4,1)
are wonderfully captured by the linear PN approximations at all
epochs.

The major exception is presented by the (3,2) mode's amplitude for equal mass systems
which shows a curious, but definite nonlinear dependence on its effective spin
\emph{at all studied epochs}. This behavior is curious for two reasons: The
PN signature is the strongest for the other three modes at mass ratio $q=1$
and all epochs (cf. Fig.~\ref{fig:aligned_plots_PN_mass_ratio_series}, first column).
Further, it is the only case (i.e. the only mode for both the aligned and non-spinning
simulations) where the PN approximation seems to already fail at $\tau=-100\, M$.
It shows that the (3,2) modes takes a special place amongst the $\ell+m=\text{odd}$
modes with $\ell \leq 4$. This is captured by its even azimuthal number $m=2$ whereas
the other three are odd $m$ modes. Thus, the (3,2) mode is the most interesting mode
amongst all the weaker $\ell\neq m$ modes to study deviations from the PN
signature: It is the strongest of these modes and thus the easiest to detect, it does
not vanish for non-spinning, equal mass systems, and it is affected by spin effects
where it can capture departures from PN theory well into the inspiral-regime.

\section{Conclusions}
In this paper we have provided a comparison of the amplitudes of spherical harmonic modes 
of gravitational waves from merging binary black holes computed using the leading order PN
approximation with those obtained from numerical relativity simulations.

The post-Newtonian
approximation is based on the point-particle description of the two-body problem in general
relativity. It is a good approximation when the two bodies are far from each other (i.e.,
their distance of separation $r$ is far greater than the scale of the horizon $R_s\sim 2GM/c^2$ 
of the component masses), but expected to breakdown when the two bodies are close to 
coalescence $r \sim \mbox{few}\times 2GM/c^2.$ While the post-Newtonian approximation is
now known to a high order in the expansion parameter $v/c = \sqrt{GM/c^2r},$ it is not expected to
capture the strong field dynamics of the theory close to merger.

Numerical relativity simulations, on the contrary, are exact solutions to Einstein's 
equations for the two-body problem. They capture the strong field dynamics, including 
the dynamics of the common horizon and how that horizon approaches the final Kerr state. 
While these simulations can, in principle, resolve the full spectrum of modes emitted
by the binary, in practice finite resolution and numerical accuracy limit the number of
modes that can be extracted reliably to the quadrupole, octupole, and hexadecapole modes,
corresponding to spherical harmonic index of $\ell=2,3,$and $4,$ respectively.

The chief finding of our study is that the dependencies of these dominant mode
amplitudes on the symmetric mass ratio and the binary's spins, computed in the
leading order post-Newtonian approximation, agree remarkably well with those
extracted from numerical relativity simulations, deep into the regime where the
approximation should not have worked.
In particular, the quadrupole modes $(2, 2)$ and $(2,1)$, extracted from
numerical relativity simulations, show little departure from the leading order
post-Newtonian expression throughout the inspiral and merger. This is also true
for the $(3,3)$ and the $(4,4)$ modes. This implies that the dominant multipole
structure of the system remains frozen as determined by the point-particle
approximation.  All the modes begin to show significant departure from
post-Newtonian description in the quasi-normal mode regime, $\sim 10GM/c^3$
after the waveform reaches its peak amplitude.
\\
\indent The weaker modes with $\ell\!=\!3,4$, $m\!\neq\!\ell$ modes also agree
with the leading order post-Newtonian expressions when the system is $\sim 100
GM/c^3$ away from coalescence, but begin to show significant departure from the
leading order post-Newtonian behavior well-before the epoch when the waveform
reaches its peak amplitude. In other words, the point-particle approximation of
post-Newtonian theory to the leading order is no longer adequate in describing
the behavior of the amplitude of these modes. It is for this reason that we
conclude that these weaker modes are affected far more by the strong field
regime of the binary evolution than the stronger modes $(2,2), (2,1), (3,3)$
and $(4,4)$. 
\\
\indent It is well known that the $(3,2)$ spherical harmonic mode is a mixture
of several spheroidal harmonic modes, which causes it to decay
non-monotonically in the ringdown regime of the signal
\cite{Kelly:2012nd,London:2014cma,London2018}. While this is true, the new
insight from our study is that we can exploit the leading order post-Newtonian
expressions in any analytical modeling of the mode amplitudes. Furthermore, we
believe that understanding the multipole structure of the common horizon could
provide further insight into why certain modes are affected far more by the
strong field dynamics than others.

\acknowledgements
We thank Abhay Ashtekar, Michael Boyle, Mark Hannam, Lionel London, Sean McWilliams and Leo Stein 
for helpful comments on the manuscript.  KGA and BSS acknowledge the support by the Indo-US Science 
and Technology Forum through the Indo-US {\em Centre for the Exploration of Extreme Gravity}, 
grant IUSSTF/JC-029/2016. SB and BSS are supported in part by NSF grants PHY-1836779, AST-1716394 
and AST-1708146.  KGA is partially support by a grant from Infosys Foundation. KGA also acknowledge 
partial support by the grant EMR/2016/005594. Computing resources for this project were provided 
by The Pennsylvania State University.  This document has LIGO preprint number {\tt LIGO-P1800367}.  

\appendix
%
\section{Alternative representation of the results for aligned spin fits} \label{app:alternative_mass_ratio_series}
Fig.~\ref{fig:aligned_plots_PN_mass_ratio_series} presents the same information
as Fig.~\ref{fig:aligned_plots_PN_time_series}, but with the roles of the time
$\tau$ and mass ratio $q$ flipped in the figure. This presentation allows a
more streamlined look at how each mode behaves as function of time for a given
mass ratio, thus making very evident, how strongly the (3,2) modes deviates
from the leading order post-Newtonian approximation for mass ratio $q=1$.
\section{Numerical relativity simulations from the SXS project} \label{app:SXS_simulations}
For each numerical resolution, the SXS waveform catalog provides a
metadata file with information about the specifics of the run as well
as the gravitational waveforms decomposed into spherical harmonics for
both the Newman-Penrose scalar $\Psi_{4}$ and the gravitational wave strain $h$.  Our
analysis focuses on the latter and was conducted with the data
contained in the files `rhOverM\_Asymptotic\_GeometricUnits\_CoM.h5',
which provide the spherical harmonic modes of $h$, at the outermost extraction
radius, and extrapolated to
asymptotic null infinity.  Furthermore, the data in these files are
corrected for mode mixing that can arise if initial transients during
start of the evolution induce a motion of the center of mass of the
binary black hole~\cite{Boyle:2015nqa,Ossokine:2015yla}. The retarded
time-coordinate is corrected for gravitational redshift
effects~\cite{Boyle:2009vi}.
\\
These HDF5 files structure the data into four groups
containing the same 77 datasets, but for different extrapolation
orders $N=2,3,4$, as well as the outermost extraction radius. The
datasets store the simulation output as a time series of real and
imaginary components of the coefficients $h_{\ell m}$ in the expansion in
spherical harmonics with spin-weight $s=-2$
\eqref{eq:polarisations_and_modes} of the gravitational wave strain $h$ for all 77
modes with $\ell=2,\dots,8$, $m=-\ell,\dots,\ell$.
In order to put errors on the numerical data we restricted our analysis to the
43 non-spinning and 121 aligned spinning simulations that are provided at a
minimum of two different resolution levels, see Tables \ref{tab:simulations_ns}
and \ref{tab:simulations_as}. Further, we restrict our analysis to the
outermost extraction radius which yields the most accurate numerical results
for the merger and ringdown regimes.
\begin{table}[t]
        \begin{tabular}{|c|c|c|}
        \hline
        SXS Id & $q$ & Resolutions \\ 
        \hline
         $2$ & $1.00$ & $4$, $5$, $6$ \\ 
        \hline
         $180$ & $1.00$ & $2$, $3$, $4$ \\ 
        \hline
         $198$ & $1.20$ & $3$, $4$, $5$ \\ 
        \hline
         $7$ & $1.50$ & $4$, $5$ \\ 
        \hline
         $8$ & $1.50$ & $4$, $5$ \\ 
        \hline
         $194$ & $1.52$ & $2$, $3$ \\ 
        \hline
         $169$ & $2.00$ & $3$, $4$, $5$ \\ 
        \hline
         $184$ & $2.00$ & $2$, $3$, $4$ \\ 
        \hline
         $201$ & $2.32$ & $1$, $2$, $3$ \\ 
        \hline
         $259$ & $2.50$ & $3$, $4$, $5$ \\ 
        \hline
         $191$ & $2.51$ & $2$, $3$ \\ 
        \hline
         $30$ & $3.00$ & $3$, $4$, $5$ \\ 
        \hline
         $168$ & $3.00$ & $3$, $4$, $5$ \\ 
        \hline
         $183$ & $3.00$ & $2$, $3$, $4$ \\ 
        \hline
         $200$ & $3.27$ & $1$, $2$, $3$ \\ 
        \hline
         $193$ & $3.50$ & $2$, $3$ \\ 
        \hline
         $294$ & $3.50$ & $3$, $4$ \\ 
        \hline
         $182$ & $4.00$ & $2$, $3$, $4$ \\ 
        \hline
         $190$ & $4.50$ & $2$, $3$ \\ 
        \hline
         $54$ & $5.00$ & $3$, $4$, $5$ \\ 
        \hline
         $56$ & $5.00$ & $3$, $4$, $5$ \\ 
        \hline
         $107$ & $5.00$ & $3$, $4$, $5$ \\ 
        \hline
        \end{tabular}
        \hspace*{0.5em}
        \begin{tabular}{|c|c|c|}
        \hline
        SXS Id & $q$ & Resolutions \\ 
        \hline
        $113$ & $5.00$ & $3$, $4$, $5$ \\ 
        \hline
         $187$ & $5.04$ & $1$, $2$, $3$ \\ 
        \hline
         $296$ & $5.50$ & $3$, $4$, $5$ \\ 
        \hline
         $197$ & $5.52$ & $2$, $3$ \\ 
        \hline
         $181$ & $6.00$ & $3$, $4$ \\ 
        \hline
         $297$ & $6.50$ & $3$, $4$, $5$ \\ 
        \hline
         $192$ & $6.58$ & $2$, $3$ \\ 
        \hline
         $298$ & $7.00$ & $3$, $4$, $5$ \\ 
        \hline
         $188$ & $7.19$ & $1$, $2$, $3$ \\ 
        \hline
         $299$ & $7.50$ & $3$, $4$, $5$ \\ 
        \hline
         $195$ & $7.76$ & $2$, $3$ \\ 
        \hline
         $63$ & $8.00$ & $3$, $4$, $5$ \\ 
        \hline
         $186$ & $8.27$ & $1$, $2$, $3$ \\ 
        \hline
         $300$ & $8.50$ & $3$, $4$, $5$ \\ 
        \hline
         $199$ & $8.73$ & $2$, $3$ \\ 
        \hline
         $301$ & $9.00$ & $3$, $4$, $5$ \\ 
        \hline
         $189$ & $9.17$ & $2$, $3$ \\ 
        \hline
         $302$ & $9.50$ & $3$, $4$, $5$ \\ 
        \hline
         $196$ & $9.66$ & $2$, $3$ \\ 
        \hline
         $185$ & $9.99$ & $1$, $2$, $3$ \\ 
        \hline
         $303$ & $10.00$ & $3$, $4$, $5$ \\ 
        \hline
        \multicolumn{3}{c}{}\\
        \end{tabular}
    \caption{List of 43 SXS simulations for initially non-spinning binary black holes, showing the SXS simulation ID, the mass ratio $q$, and the numerical resolutions used for our analysis.}
\label{tab:simulations_ns}
\end{table} 
\begin{figure*}
    \includegraphics[width=0.95\textwidth]{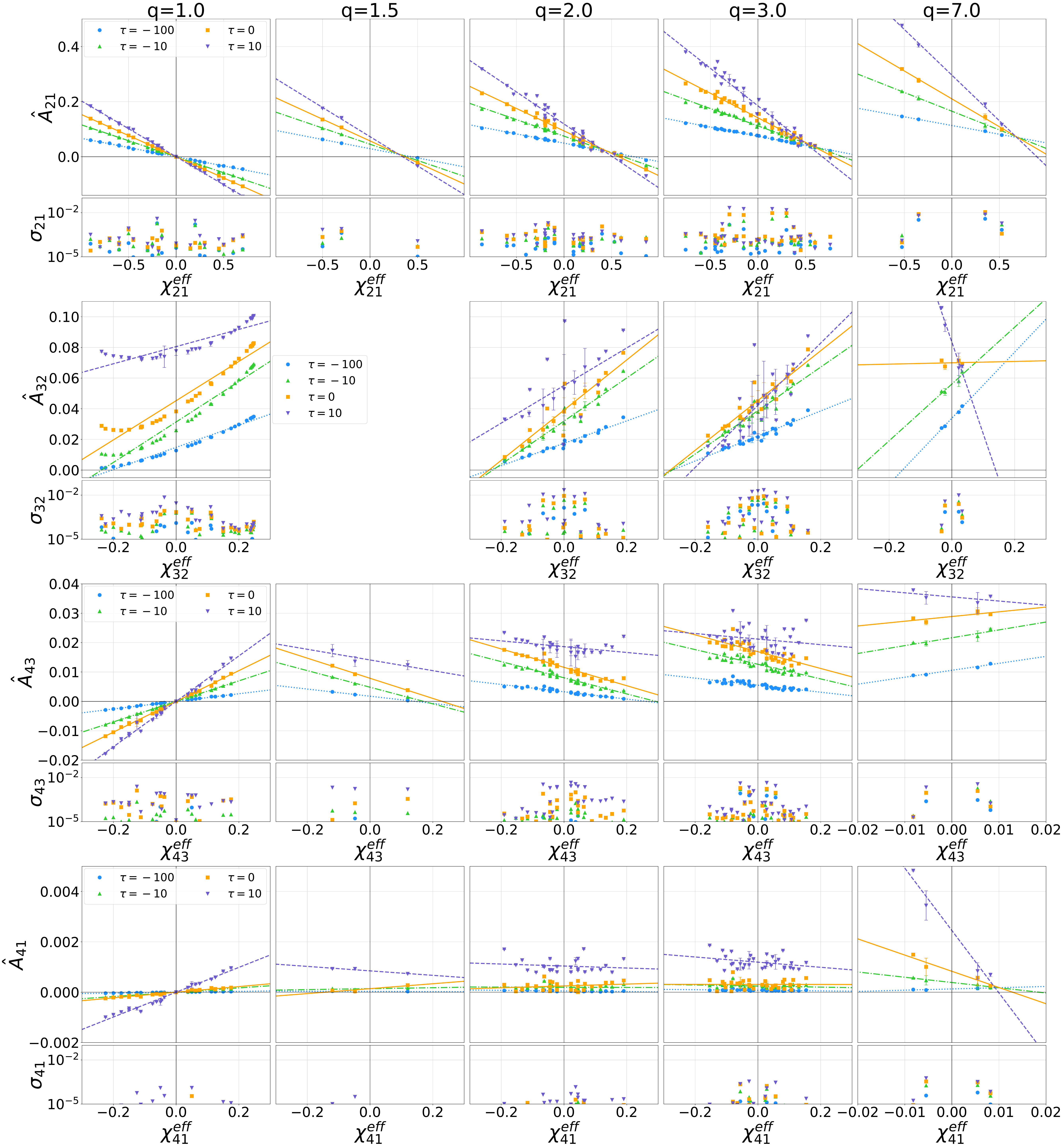}
    \caption{The same as Fig.~\ref{fig:aligned_plots_PN_time_series}, albeit the columns now represent the five mass ratios $q=1,1.5,2,3,7$ and each plot contains the data and fits at the four different times $\frac{\tau}{M}=-100,-10,0,10$. The $\chi_{lm}^{\rm eff}$ are defined in Eqs.~(\ref{eq:effective_PN_spins}). This presentaion shows more clearly how well the data is captured by the PN inspired fits (\emph{lines}) in the case of the (2,1) mode for $q=1,7$ and how the data (\emph{data points from SXS}) slightly scatters around fit lines for $q=2,3$. Similarly, this presentation makes it much clearer that fits cannot capture the amplitudes of the (3,2) mode for $q=1$, even though $q=7$ seem to be fine. Further, it shows beautifully that these modes gain in importance as time advances as well as for increasing mass ratios.}
    \label{fig:aligned_plots_PN_mass_ratio_series}
\end{figure*}
\begin{table*}
        \begin{tabular}{|c|c|r|r|c|}
        \hline
        Id & $q$ & $\chi_1$ & $\chi_2$ & Res. \\ 
        \hline
         $4$ & $1.00$ & $-0.50$ & $0.00$ & $5$, $6$ \\ 
        \hline
         $5$ & $1.00$ & $0.50$ & $0.00$ & $4$, $5$ \\ 
        \hline
         $148$ & $1.00$ & $-0.44$ & $-0.44$ & $4$, $5$ \\ 
        \hline
         $149$ & $1.00$ & $-0.20$ & $-0.20$ & $3$, $4$ \\ 
        \hline
         $150$ & $1.00$ & $0.20$ & $0.20$ & $3$, $4$ \\ 
        \hline
         $151$ & $1.00$ & $-0.60$ & $-0.60$ & $3$, $4$ \\ 
        \hline
         $152$ & $1.00$ & $0.60$ & $0.60$ & $3$, $4$ \\ 
        \hline
         $154$ & $1.00$ & $-0.80$ & $-0.80$ & $3$, $4$ \\ 
        \hline
         $155$ & $1.00$ & $0.80$ & $0.80$ & $3$, $4$ \\ 
        \hline
         $156$ & $1.00$ & $-0.95$ & $-0.95$ & $4$, $5$ \\ 
        \hline
         $157$ & $1.00$ & $0.95$ & $0.95$ & $3$, $4$ \\ 
        \hline
         $158$ & $1.00$ & $0.97$ & $0.97$ & $5$, $6$ \\ 
        \hline
         $159$ & $1.00$ & $-0.90$ & $-0.90$ & $3$, $4$ \\ 
        \hline
         $160$ & $1.00$ & $0.90$ & $0.90$ & $3$, $4$ \\ 
        \hline
         $170$ & $1.00$ & $0.44$ & $0.44$ & $5$, $6$ \\ 
        \hline
         $171$ & $1.00$ & $-0.44$ & $-0.44$ & $5$, $6$ \\ 
        \hline
         $172$ & $1.00$ & $0.98$ & $0.98$ & $3$, $4$ \\ 
        \hline
         $175$ & $1.00$ & $0.75$ & $0.75$ & $2$, $3$ \\ 
        \hline
         $176$ & $1.00$ & $0.96$ & $0.96$ & $3$, $4$ \\ 
        \hline
         $177$ & $1.00$ & $0.99$ & $0.99$ & $3$, $4$ \\ 
        \hline
         $178$ & $1.00$ & $0.99$ & $0.99$ & $4$, $5$ \\ 
        \hline
         $209$ & $1.00$ & $-0.90$ & $-0.50$ & $4$, $5$ \\ 
        \hline
         $210$ & $1.00$ & $-0.90$ & $0.00$ & $4$, $5$ \\ 
        \hline
         $211$ & $1.00$ & $-0.90$ & $0.90$ & $4$, $5$ \\ 
        \hline
         $212$ & $1.00$ & $-0.80$ & $-0.80$ & $4$, $5$ \\ 
        \hline
         $213$ & $1.00$ & $-0.80$ & $0.80$ & $4$, $5$ \\ 
        \hline
         $214$ & $1.00$ & $-0.62$ & $-0.25$ & $4$, $5$ \\ 
        \hline
         $215$ & $1.00$ & $-0.60$ & $-0.60$ & $4$, $5$ \\ 
        \hline
         $216$ & $1.00$ & $-0.60$ & $0.00$ & $4$, $5$ \\ 
        \hline
         $217$ & $1.00$ & $-0.60$ & $0.60$ & $4$, $5$ \\ 
        \hline
         $218$ & $1.00$ & $-0.50$ & $0.50$ & $4$, $5$ \\ 
        \hline
        \end{tabular}
\hspace*{0.5em}
        \begin{tabular}{|c|c|r|r|c|}
          \hline
          Id & $q$ & $\chi_1$ & $\chi_2$ & Res. \\ 
          \hline
         $219$ & $1.00$ & $-0.50$ & $0.90$ & $4$, $5$ \\ 
        \hline
         $220$ & $1.00$ & $-0.40$ & $-0.80$ & $4$, $5$ \\ 
        \hline
         $221$ & $1.00$ & $-0.40$ & $0.80$ & $4$, $5$ \\ 
        \hline
         $222$ & $1.00$ & $-0.30$ & $0.00$ & $4$, $5$ \\ 
        \hline
         $223$ & $1.00$ & $0.30$ & $0.00$ & $4$, $5$ \\ 
        \hline
         $224$ & $1.00$ & $0.40$ & $-0.80$ & $4$, $5$ \\ 
        \hline
         $225$ & $1.00$ & $0.40$ & $0.80$ & $4$, $5$ \\ 
        \hline
         $226$ & $1.00$ & $0.50$ & $-0.90$ & $4$, $5$ \\ 
        \hline
         $227$ & $1.00$ & $0.60$ & $0.00$ & $4$, $5$ \\ 
        \hline
         $228$ & $1.00$ & $0.60$ & $0.60$ & $4$, $5$ \\ 
        \hline
         $229$ & $1.00$ & $0.65$ & $0.25$ & $4$, $5$ \\ 
        \hline
         $230$ & $1.00$ & $0.80$ & $0.80$ & $4$, $5$ \\ 
        \hline
         $231$ & $1.00$ & $0.90$ & $0.00$ & $4$, $5$ \\ 
        \hline
         $232$ & $1.00$ & $0.90$ & $0.50$ & $4$, $5$ \\ 
        \hline
         $304$ & $1.00$ & $0.50$ & $-0.50$ & $3$, $4$ \\ 
        \hline
         $12$ & $1.50$ & $-0.50$ & $0.00$ & $4$, $5$ \\ 
        \hline
         $14$ & $1.50$ & $-0.50$ & $0.00$ & $4$, $5$ \\ 
        \hline
         $16$ & $1.50$ & $-0.50$ & $0.00$ & $5$, $6$ \\ 
        \hline
         $19$ & $1.50$ & $-0.50$ & $0.50$ & $4$, $5$ \\ 
        \hline
         $25$ & $1.50$ & $0.50$ & $-0.50$ & $4$, $5$ \\ 
        \hline
         $162$ & $2.00$ & $0.60$ & $0.00$ & $3$, $4$ \\ 
        \hline
         $233$ & $2.00$ & $-0.87$ & $0.85$ & $4$, $5$ \\ 
        \hline
         $234$ & $2.00$ & $-0.85$ & $-0.85$ & $4$, $5$ \\ 
        \hline
         $235$ & $2.00$ & $-0.60$ & $-0.60$ & $4$, $5$ \\ 
        \hline
         $236$ & $2.00$ & $-0.60$ & $0.00$ & $4$, $5$ \\ 
        \hline
         $237$ & $2.00$ & $-0.60$ & $0.60$ & $4$, $5$ \\ 
        \hline
         $238$ & $2.00$ & $-0.50$ & $-0.50$ & $4$, $5$ \\ 
        \hline
         $239$ & $2.00$ & $-0.37$ & $0.85$ & $4$, $5$ \\ 
        \hline
         $240$ & $2.00$ & $-0.30$ & $-0.30$ & $4$, $5$ \\ 
        \hline
         $241$ & $2.00$ & $-0.30$ & $0.00$ & $4$, $5$ \\ 
        \hline
         $242$ & $2.00$ & $-0.30$ & $0.30$ & $4$, $5$ \\ 
        \hline
        \end{tabular}
\hspace*{0.5em}
        \begin{tabular}{|c|c|r|r|c|}
        \hline
        Id &  $q$ & $\chi_1$ & $\chi_2$ & Res. \\ 
        \hline
         $243$ & $2.00$ & $-0.13$ & $-0.85$ & $4$, $5$ \\ 
        \hline
         $244$ & $2.00$ & $0.00$ & $-0.60$ & $4$, $5$ \\ 
        \hline
         $245$ & $2.00$ & $0.00$ & $-0.30$ & $4$, $5$ \\ 
        \hline
         $246$ & $2.00$ & $0.00$ & $0.30$ & $4$, $5$ \\ 
        \hline
         $247$ & $2.00$ & $0.00$ & $0.60$ & $4$, $5$ \\ 
        \hline
         $248$ & $2.00$ & $0.13$ & $0.85$ & $4$, $5$ \\ 
        \hline
         $249$ & $2.00$ & $0.30$ & $-0.30$ & $4$, $5$ \\ 
        \hline
         $250$ & $2.00$ & $0.30$ & $0.00$ & $4$, $5$ \\ 
        \hline
         $251$ & $2.00$ & $0.30$ & $0.30$ & $4$, $5$ \\ 
        \hline
         $252$ & $2.00$ & $0.37$ & $-0.85$ & $4$, $5$ \\ 
        \hline
         $253$ & $2.00$ & $0.50$ & $0.50$ & $4$, $5$ \\ 
        \hline
         $254$ & $2.00$ & $0.60$ & $-0.60$ & $4$, $5$ \\ 
        \hline
         $255$ & $2.00$ & $0.60$ & $0.00$ & $4$, $5$ \\ 
        \hline
         $256$ & $2.00$ & $0.60$ & $0.60$ & $4$, $5$ \\ 
        \hline
         $257$ & $2.00$ & $0.85$ & $0.85$ & $4$, $5$ \\ 
        \hline
         $258$ & $2.00$ & $0.87$ & $-0.85$ & $4$, $5$ \\ 
        \hline
         $31$ & $3.00$ & $0.50$ & $0.00$ & $4$, $5$ \\ 
        \hline
         $36$ & $3.00$ & $-0.50$ & $0.00$ & $5$, $6$ \\ 
        \hline
         $174$ & $3.00$ & $0.50$ & $0.00$ & $5$, $6$ \\ 
        \hline
         $260$ & $3.00$ & $-0.85$ & $-0.85$ & $4$, $5$ \\ 
        \hline
         $261$ & $3.00$ & $-0.73$ & $0.85$ & $4$, $5$ \\ 
        \hline
         $262$ & $3.00$ & $-0.60$ & $0.00$ & $4$, $5$ \\ 
        \hline
         $263$ & $3.00$ & $-0.60$ & $0.60$ & $4$, $5$ \\ 
        \hline
         $264$ & $3.00$ & $-0.60$ & $-0.60$ & $4$, $5$ \\ 
        \hline
         $265$ & $3.00$ & $-0.60$ & $-0.40$ & $4$, $5$ \\ 
        \hline
         $266$ & $3.00$ & $-0.60$ & $0.40$ & $4$, $5$ \\ 
        \hline
         $267$ & $3.00$ & $-0.50$ & $-0.50$ & $4$, $5$ \\ 
        \hline
         $268$ & $3.00$ & $-0.40$ & $-0.60$ & $4$, $5$ \\ 
        \hline
         $269$ & $3.00$ & $-0.40$ & $0.60$ & $4$, $5$ \\ 
        \hline
         $270$ & $3.00$ & $-0.30$ & $-0.30$ & $4$, $5$ \\ 
        \hline
         $271$ & $3.00$ & $-0.30$ & $0.00$ & $4$, $5$ \\ 
        \hline
        \end{tabular}
\hspace*{0.5em}
        \begin{tabular}{|c|c|r|r|c|}
          \hline
          Id & $q$ & $\chi_1$ & $\chi_2$ & Res. \\ 
          \hline
         $272$ & $3.00$ & $-0.30$ & $0.30$ & $4$, $5$ \\ 
        \hline
         $273$ & $3.00$ & $-0.27$ & $-0.85$ & $4$, $5$ \\ 
        \hline
         $274$ & $3.00$ & $-0.23$ & $0.85$ & $4$, $5$ \\ 
        \hline
         $275$ & $3.00$ & $0.00$ & $-0.60$ & $4$, $5$ \\ 
        \hline
         $276$ & $3.00$ & $0.00$ & $-0.30$ & $4$, $5$ \\ 
        \hline
         $277$ & $3.00$ & $0.00$ & $0.30$ & $4$, $5$ \\ 
        \hline
         $278$ & $3.00$ & $0.00$ & $0.60$ & $4$, $5$ \\ 
        \hline
         $279$ & $3.00$ & $0.23$ & $-0.85$ & $4$, $5$ \\ 
        \hline
         $280$ & $3.00$ & $0.27$ & $0.85$ & $4$, $5$ \\ 
        \hline
         $281$ & $3.00$ & $0.30$ & $-0.30$ & $4$, $5$ \\ 
        \hline
         $282$ & $3.00$ & $0.30$ & $0.00$ & $4$, $5$ \\ 
        \hline
         $283$ & $3.00$ & $0.30$ & $0.30$ & $4$, $5$ \\ 
        \hline
         $284$ & $3.00$ & $0.40$ & $-0.60$ & $4$, $5$ \\ 
        \hline
         $285$ & $3.00$ & $0.40$ & $0.60$ & $4$, $5$ \\ 
        \hline
         $286$ & $3.00$ & $0.50$ & $0.50$ & $4$, $5$ \\ 
        \hline
         $287$ & $3.00$ & $0.60$ & $-0.60$ & $4$, $5$ \\ 
        \hline
         $288$ & $3.00$ & $0.60$ & $-0.40$ & $4$, $5$ \\ 
        \hline
         $289$ & $3.00$ & $0.60$ & $0.00$ & $4$, $5$ \\ 
        \hline
         $290$ & $3.00$ & $0.60$ & $0.40$ & $4$, $5$ \\ 
        \hline
         $291$ & $3.00$ & $0.60$ & $0.60$ & $4$, $5$ \\ 
        \hline
         $292$ & $3.00$ & $0.73$ & $-0.85$ & $4$, $5$ \\ 
        \hline
         $293$ & $3.00$ & $0.85$ & $0.85$ & $4$, $5$ \\ 
        \hline
         $202$ & $7.00$ & $0.60$ & $0.00$ & $3$, $4$ \\ 
        \hline
         $203$ & $7.00$ & $0.40$ & $0.00$ & $2$, $3$ \\ 
        \hline
         $204$ & $7.00$ & $0.40$ & $0.00$ & $2$, $3$ \\ 
        \hline
         $205$ & $7.00$ & $-0.40$ & $0.00$ & $2$, $3$ \\ 
        \hline
         $206$ & $7.00$ & $-0.40$ & $0.00$ & $2$, $3$ \\ 
        \hline
         $207$ & $7.00$ & $-0.60$ & $0.00$ & $3$, $4$ \\ 
        \hline
        \multicolumn{3}{c}{}\\
        \multicolumn{3}{c}{}\\
        \multicolumn{3}{c}{}\\
        \end{tabular}
        \caption{List of 121 SXS simulations for aligned-spin binary black holes, showing the SXS simulation ID, the mass ratio $q$, the spins represented by $\chi_{1,2}$ via $\vec\chi_{1,2}=\chi_{1,2}\hat L$, and the numerical resolutions used for our analysis.}
\label{tab:simulations_as}
\end{table*}
\FloatBarrier
\bibliography{refs}
\bibliographystyle{apsrev}	
\end{document}